\documentclass[11pt,a4paper]{article}
\pdfoutput=1
\usepackage[pdftex]{graphicx}
\usepackage{jheppub}
\usepackage{bm}
\usepackage{amsmath} 
\usepackage{amsthm}
\usepackage{amssymb} 
\usepackage{MnSymbol}
\usepackage{hyperref}
\usepackage{xfrac}
\usepackage{subfig}

\newcommand{\be}{\begin{equation}}
\newcommand{\ee}{\end{equation}}
\newcommand{\bea}{\begin{eqnarray}}
\newcommand{\eea}{\end{eqnarray}}

\def\part#1#2{\frac{\partial#1}{\partial#2}}

\def\rhox{\rho_\times}
\def\CF{\mathcal{F}}
\def\CH{\mathcal{H}}
\def\CL{\mathcal{L}}
\def\CM{\mathcal{M}}
\def\CN{\mathcal{N}}
\def\CO{\mathcal{O}}
\def\CZ{\mathcal{Z}}

\begin{document}

\title{Wilsonian renormalisation and the exact cut-off scale from holographic duality}
\author{Sa\v{s}o Grozdanov}
\affiliation{Rudolf Peierls Centre for Theoretical Physics, University of Oxford, \\ 1 Keble Road, Oxford OX1 3NP, United Kingdom}
\emailAdd{s.grozdanov1@physics.ox.ac.uk}

\date{\today}

\begin{abstract}{
We propose a method for determining the exact correspondence between the Wilsonian cut-off scale on the boundary and its holographically dual bulk theory. We systematically construct the multi-trace Wilsonian effective action from holographic renormalisation and evolve it by integrating out the asymptotically Anti-de Sitter bulk geometry with scalar probes. The Wilsonian nature of the effective action is shown by proving that it must be either double-trace, closing in on itself under successive integrations, or have an infinite series of multi-trace terms. Focusing on composite scalar operator renormalisation, we relate the Callan-Symanzik equation, the flow of the scalar anomalous dimension and the multi-trace beta functions to their dual RG flows in the bulk. Establishing physical renormalisation conditions on the behaviour of the large-$N$ anomalous dimension then enables us to extract the energy scales. Examples of pure AdS, GPPZ flow, black brane in AdS, M2 and M5 branes are studied before we generalise our results to arbitrary numbers of mass and thermal deformations of an ultra-violet CFT. Relations between the undeformed Wilsonian cut-off, deformation scales and the deformed Wilsonian cut-off are discussed, as is phenomenology of each considered background. We see how a mass gap, the emergent infra-red CFT scaling, etc. arise in different effective theories. We also argue that these results can have alternative interpretations through the flow of the conformal anomaly or the Ricci scalar curvature of boundary branes. They show consistency with the c-theorem.}
\end{abstract}


\begin{flushright}OUTP-11-58P \end{flushright}
\maketitle
\flushbottom

\section{Introduction}

Beyond its fundamental conceptual importance to physics, the holographic duality or AdS/CFT correspondence \cite{Maldacena:1997re,Gubser:1998bc,Witten:1998qj} has provided us with a very useful tool for the study of strongly coupled field theories. The correspondence enables us to calculate field theory predictions directly from their weakly coupled classical gravity duals. The string theory, or gravity, side of the duality is formulated in a bulk spacetime with an extra radial dimension compared to its holographic field theory dual, which lives on the bulk's boundary. The radial coordinate can be understood as an energy scale of the dual theory \cite{Maldacena:1997re,Susskind:1998dq,Peet:1998wn}. In an asymptotically Anti-de Sitter spacetime, the metric tensor diverges at the boundary, which corresponds to a UV divergence of the field theory. This further establishes the IR/UV correspondence between the two dual theories \cite{Susskind:1998dq,Peet:1998wn}.    

Details of the strongly coupled field theories dual to theories in the bulk are usually unknown. An interesting question in its own right therefore becomes even more crucial for further understanding of the duality: how can we precisely understand renormalisation group flows of the boundary theory by exploiting the duality between its energy scale and the radial coordinate? Since the early days of the correspondence numerous papers have explored this question, with some among the earlier ones: \cite{Akhmedov:1998vf,Girardello:1998pd,Distler:1998gb,Balasubramanian:1999jd,de Boer:1999xf,Balasubramanian:1998de,Freedman:1999gp,deBoer:2000cz}. 

A conformal field theory at the AdS infinity can naturally be interpreted as a UV fixed point with its cut-off taken to infinity, without the theory running into problems such as a Landau pole. We would then like to know how to formulate the Wilsonian procedure of integrating out the UV degrees of freedom and flowing towards the IR \cite{Wilson:1973jj,Wegner:1972ih,Wilson:1993dy,Polchinski:1983gv,Polonyi:2001se}. This can, in fact, be done on the gravity side, as shown by Faulkner, Liu and Rangamani in \cite{Faulkner:2010jy} and Heemskerk and Polchinski in \cite{Heemskerk:2010hk}, in a way which is completely analogous to the Wilsonian renormalisation group. The procedure is based on integrating out shells of geometry along the radial direction, while keeping the overall path integral fixed. Other recent developments on the subject include \cite{Sin:2011yh,Harlow:2011ke,Fan:2011wm,Akhmedov:2010mz,Radicevic:2011py,Elander:2011vh,Laia:2011wf,Bredberg:2010ky,Lee:2009ij,Lee:2010ub,Paulos:2011zu,Kuperstein:2011fn}. 

It is the goal of this paper to further develop the correspondence between the Wilsonian picture and the bulk integration of geometry. In particular, we wish to propose a systematic procedure for finding an \textit{exact dictionary between the hard Wilsonian cut-off scale on the field theory side and its dual bulk description}. We also wish to show that the \textit{effective Wilsonian action must either include an infinite set of multi-trace operators or close in on its double-trace sector under the renormalisation group flow.} This would provide additional convincing evidence that it is possible to construct, directly from holography, a renormalisation group procedure of the boundary field theory which is fully Wilsonian. Understanding how different energy scales in a field theory, and their mixing, carry over from the dual geometric picture is important for making holography more precise. Furthermore, it is of considerable interest to the model-building of condensed matter and particle theory systems with holographic duals. Most realistic systems do not exhibit conformal, or scale, invariance. It is therefore useful to understand how various UV conformal theories can be deformed to reproduce some desired phenomenological behaviour by their effective IR theories. Once we have found such a bulk theory, all calculations could then be performed at the conformal point where holography is well understood. Physical predictions, such as the renormalised correlation functions remain invariant under renormalisation group transformations, thus providing us with a source of descriptions of IR phenomena by IR effective theories with UV completions.

We focus on asymptotically Anti-de Sitter bulks with propagating scalar fields. Specifically, we wish to understand the proportionality between the energy scale and the radial coordinate. Once this is understood for conformal theories, such as the $\CN=4$ super Yang-Mills, we study the same theory with mass, thermal and density deformations to see how they alter the RG running of the Wilsonian scale. Further cases are also studied of the near-extremal M2 and M5 branes at finite temperature. Finally, we generalise the procedure to include an arbitrary number of mass deformations and Lorentz-invariance breaking thermal deformations, with the latter being induced by horizons of black branes. We will also analyse the phenomenology of each studied example and extract various physical interpretations of theories from the flows of the RG scales. Among them will be the indications for the existence of a mass gap in the confining GPPZ case, a mass gap and the emergent infra-red CFT scaling in duals of black branes in AdS, the mixing of different energy scales, etc.   

In order to develop the procedure of extracting the dependence of the cut-off on the bulk physics, we will establish a systematic way of finding the bare boundary action, which will run under the RG transformations of integrating out the bulk. We first use the fact that in order to have a well-defined and finite holographic description at the AdS infinity, we need to regulate and renormalise the boundary action. This is done using holographic renormalisation \cite{de Haro:2000xn,Skenderis:2002wp,Bianchi:2001kw,Skenderis:2008dg}, whereby counter-terms are introduced to exactly cancel off divergences in the limit of the AdS infinity. We use a combination of the regularised action and holographic counter-terms to write the bare action, and then permit various terms in it to run. The procedure yields the bare effective boundary action with extra terms, which in the Wilsonian RG come from the structure of counter-terms. This agrees with the expectation that the additional effective terms must be invariant under the full symmetries of the field theory. But these are precisely the isometries of the bulk under which the counter-terms must transform. The procedure established in \cite{Faulkner:2010jy} can then be used to evolve the effective action under the RG flow. Through a careful definition of the wavefunction renormalisation, the double-trace coupling can be related to the anomalous dimension of scalar operators in large-$N$ theories. The role of multi-trace deformations in AdS/CFT was studied, among other papers, in \cite{Witten:2001ua,Akhmedov:2002gq,Mueck:2002gm,Minces:2002wp,Vecchi:2010dd,Hartman:2006dy}. From the large-$N$ field theory point of view, these works relate to \cite{Pomoni:2008de,Vecchi:2010jz}, for example.  

Once we have established how the Callan-Symanzik equation and the flow of the anomalous dimension of the field theory operators translate to the RG equations in the bulk, we can then analyse them in various scenarios. It is the flow of the anomalous dimension, proportional to the double-trace beta function, that allows us to find the exact dependence of the Wilsonian cut-off scale on quantities describing the physics in the bulk. This is done through a set of physical renormalisation conditions that the wavefunction renormalisation, and hence the anomalous dimension, must satisfy in any field theory. They tell us, given some physical operator momentum we wish to probe, where the flow will terminate as deep in the bulk as possible. The position can then be directly translated into the Wilsonian cut-off, which cannot be lowered below some physical momentum scale of interest. We will also see how these conditions can be shown to be consistent with the c-theorem \cite{Zam:1986,Cardy:1988,Komargodski:2011vj,Freedman:1999gp,Myers:2010xs,Gubser:2002vv,Hartman:2006dy}.

The paper is structured as follows: \\
In section \ref{Sec:RG}, we set up the procedure of obtaining the Wilsonian renormalisation group from holographic renormalisation and integrating out bulk geometry for theories with bulk scalars. In \ref{Sec:RG1} we show how to construct the running bare effective boundary action from the bulk physics, by using the structure of the holographic counter-terms. In \ref{Sec:RG2} we use the work of \cite{Faulkner:2010jy} to derive the full set of renormalisation group equations describing the flow of the effective action. In \ref{Sec:RG3} we use a definition of the wavefunction renormalisation to establish the connection between the bulk construction and the Callan-Symanzik equation on the field theory side of the duality. This results in an interpretation of the Wilsonian scalar composite operator renormalisation. We also comment on thermal scalings of operator dimensions. \\
In section \ref{Sec:AnomDim} we examine physical conditions that the anomalous dimension should obey, in order to establish boundary conditions for the differential RG equation describing its flow. By evolving the UV Wilsonian cut-off down to some operator momentum we wish to probe, or moving the brane into the bulk in the gravity picture, this allows us to find an inequality relating the physical energy-momentum and the hard cut-off. Functional dependence of the cut-off on the bulk is then examined in several cases with different bulk geometries. First we look at pure AdS with $\CN=4$ dual theory. Then we add a mass deformation (GPPZ flow), and later temperature, as well as density, by introducing a black brane into the bulk. Afterwards, we study the case of near-extremal thermal M2 and M5 branes. We also generalise the discussion to include arbitrary, simultaneous mass and thermal deformations. Relations between the hard Wilsonian cut-off and individual scales appearing in the theories are studied in detail and phenomenology of each example is discussed.  \\
Section \ref{Sec:Alter} includes a brief discussion on how our results can be alternatively interpreted as the running of the conformal anomaly or the Ricci scalar curvature of branes with dual field theories. A connection to the c-theorem is also discussed.    \\
In section \ref{Sec:MultiTrace}, we study the full infinite set of multi-trace terms in the Wilsonian effective action to provide further evidence for the Wilsonian nature of the constructed renormalisation group procedure. We show that the effective action is either quadratic, in which case it closes in on itself under the RG flow, or infinite, which is in full accordance with field theoretic expectations. We also explore the systematics for finding a formal solution of the entire set of multi-trace couplings. \\  
Finally, in section \ref{Sec:Sum}, we comment on the results of our analysis and point out some open questions.

\section{From holographic to Wilsonian renormalisation group}\label{Sec:RG}
\subsection{The effective boundary action}\label{Sec:RG1}
 
We begin our analysis by systematically constructing the Wilsonian renormalisation group from holography. We will focus on scalar operators and only consider scalar field theories in the bulk, as dictated by the AdS/CFT dictionary \cite{Gubser:1998bc,Witten:1998qj}. Throughout this work, Lorentzian bulk metrics $G_{MN}$ will have an asymptotically Anti-de Sitter boundary at radial position $r=0$ in Poincar\'{e}-like coordinates to ensure a dual field theory with a conformal UV fixed point. We use capital Latin indices for $d+1$ bulk dimensions and reserve Greek indices for induced metrics $g_{\mu\nu}$ on $d$-dimensional branes with dual field theories. The metric of the pure $AdS_{d+1}$ is 
\begin{equation}\label{AdSMet1}
ds^2 = \frac{L^2}{r^2} \left( -dt^2 + d\vec{x}^2_{d-1} +dr^2  \right).
\end{equation}
We set the AdS radius to $L=1$ throughout this work. 

Wilsonian renormalisation group analysis starts with a quantum field theory with a UV cut-off $\Lambda_0$ \cite{Wilson:1973jj,Wegner:1972ih,Wilson:1993dy,Polchinski:1983gv,Polonyi:2001se}. It is important for us to be able to take $\Lambda_0 \to \infty$ without running into problems, such as a Landau pole, and to begin the flow at a fixed point. In field theory, a momentum shell is then integrated out, leaving us with an effective Lagrangian, which includes all possible terms permitted by the symmetries. The theory also has a new hard cut-off $\Lambda_1$. On the gravity side, a dual procedure was recently proposed in \cite{Faulkner:2010jy,Heemskerk:2010hk}. It was shown that by integrating out shells of geometry along the radial coordinate, an effective theory is produced on the boundary, which had been moved further into the bulk, from $r=\rho_0$ to $\rho_1$. This adheres to the expectation that the energy scale of a field theory is inversely proportional to the radial coordinate, $\Lambda \propto 1 / r$ \cite{Susskind:1998dq,Peet:1998wn}. Having integrated out a slice between $\rho_0$, which relates to $\Lambda_0$, and $\rho_1$, we expect the lowered Wilsonian cut-off to be a function $\Lambda_1 (\rho_1)$. Determining the exact functional dependence of $\Lambda (\rho)$ is the goal of this analysis. Note that we will be using the variable $\rho$ to specify the radial position of the brane with a dual, distinguishing it from a coordinate variable $r$.    

The first question we need to address is how to systematically obtain additional terms in the effective boundary action so that we can use the RG evolution procedure of \cite{Faulkner:2010jy}. To show that this can be done by using the structure of the holographic counter-terms, we first note that all our flows start from the AdS boundary where the metric, as well as the boundary action from which dual correlation functions are extracted, diverge. It is standard to use holographic renormalisation \cite{de Haro:2000xn,Skenderis:2002wp,Bianchi:2001kw,Skenderis:2008dg} to regulate these divergences by defining the brane theory at $r=\rho_0$, dual to $\Lambda_0$, very close to the AdS infinity. 

The expression for this regulated bare boundary action,  $S_\text{B} [\rho_0] \equiv S_\text{B}^{\text{reg}} [\rho_0 ]$, is obtained from the boundary term of the gravitational bulk action $S_\text{bulk}$ that remains after using classical equations of motion. In fact, $S_\text{bulk} = - S_\text{B} [\rho_0]$, with the minus sign coming from the lower end of integration between $\rho_0$ and the horizon, or infinity. A saddle point approximation of the full path integral is allowed as we are only working with large-$N$ theories. We thus construct an action including both brane and bulk dynamics in such a way that it vanishes on-shell:
\begin{equation}\label{FullBareBulkAction}
S = S_B [\rho_0] + \!\!\!  \int \limits_{r \geq \rho_0} \!\!  d^{d+1} x \sqrt{-G} \CL \left(\Phi, \partial_M \Phi \right),
\end{equation} 
where the second term is $S_\text{bulk}$ and $\CL \left( \Phi,\partial_M \Phi \right) = -\frac{1}{2} \partial_M \Phi \partial^M \Phi - V(\Phi)$. We assume a polynomial potential $V(\Phi) = \frac{1}{2} m^2 \Phi^2 + \sum_{n=3}^\infty \frac{1}{n} b_n \Phi^n$, with the mass term kept explicit. The boundary action $S_\text{B}$ can also be viewed in the sense of \cite{Heemskerk:2010hk}, as the UV part of the bulk integral coming from the infinitesimally thin $0 \leq r \leq \rho_0$ region. Using semi-classical approximation, imagine that we are performing a path integral over $\Phi = \hat\Phi + \tilde\Phi$, where $\hat\Phi$ is the classical value and $\tilde\Phi$ a small quantum perturbation. To integrate out the region $0 \leq r \leq \rho_0$ where $\rho_0$ is infinitesimally close to $r=0$, we first integrate $S_\text{bulk}$ by parts. The setup now enables us to neglect the $(d+1)$-dimensional contribution between two boundaries. This is because the $\hat\Phi$ contribution, $\Box \hat\Phi$, vanishes by equations of motion. As for additional contributions, we assumed that $\tilde\Phi$ was very small and that the volume of space between two boundaries was infinitesimally small. At the boundaries of the bulk, the configuration space of $\Phi$ is fixed so $\tilde\Phi (0) = 0$. To a linear order in $\tilde\Phi$, and in the limit $\rho_0 \to 0$, we can cancel the classical contributions between two boundaries, $\sqrt{-G} G^{rr}\hat\Phi \partial_r \hat\Phi |_{r=\rho_0} - \sqrt{-G} G^{rr} \hat\Phi \partial_r \hat\Phi |_{r=0}$, leaving us only $ - \frac{1}{2} \int_{\rho_0} \! d^d x \sqrt{-G} G^{rr} \tilde \Phi \partial_r \tilde\Phi$. This is precisely the boundary $S_\text{B}$ we would get from the on-shell contribution of $S_\text{bulk}$. We therefore have
\begin{equation}\label{BareBdyAct}
S_{\text{B}} [\rho_0] = - \frac{1}{2}   \int \limits_{r=\rho_0}  \!\!   d^{d} x \sqrt{-G} G^{rr} \Phi \partial_r \Phi.
\end{equation} 
Treating the radial coordinate as time, we can define the bare canonical conjugate momentum as
\begin{equation}\label{BarePi}
\Pi_\text{B} \equiv \frac{\delta S_\text{B}}{\delta \Phi} = - \sqrt{-G} G^{rr} \partial_r \Phi.
\end{equation}

In the language of holographic renormalisation, we define the subtracted boundary action as
\begin{equation}\label{SubBdyAct}
S_{\text{B}}^{\text{sub}}[\rho_0] \equiv S_\text{B} [\rho_0] - S_\text{B}^{\text{c.t.}} [\rho_0],
\end{equation}
where terms in the counter-term action $S_\text{B}^{\text{c.t.}}$ are taken to exactly equal divergent pieces of $S_\text{B}$ as $\rho_0 \to 0$, resulting in a "minimal-subtraction" scheme at all momentum scales, which we will be using throughout this work. Subtracting the counter-terms from the initial $S_\text{B} [\rho_0]$ therefore makes the overall on-shell action finite in the $\rho_0 \to 0$ limit and removes all contact terms. A definition of the renormalised action naturally follows from the subtracted action via relation
\begin{equation}
S_\text{B}^{\text{ren}} \equiv \lim_{\rho_0 \to 0} S_\text{B}^{\text{sub}} [\rho_0].
\end{equation}
We further define, in analogy with $\Pi_\text{B}$, a canonical conjugate momentum
\begin{equation}\label{Pi}
\Pi \equiv \frac{\delta S_\text{B}^{\text{sub}}}{\delta \Phi},
\end{equation}
which gives, using \eqref{SubBdyAct}, a redefinition of the bare action useful for construction of the Wilsonian effective action,  
\begin{equation}\label{BareBdyAct2}
S_\text{B} [\rho_0] = \frac{1}{2} \int \limits_{r=\rho_0}   \!\!  d^d x \Pi \Phi + S_\text{B}^{\text{c.t.}} [\rho_0].
\end{equation}
The holographic counter-terms take a general form of
\begin{equation}\label{CTBdyAct}
S_\text{B}^{\text{c.t.}} [\rho_0] = - \!\!\!  \int \limits_{r=\rho_0}  \! \! \!   d^d x \sqrt{-g} \left( \frac{\Delta_{-}}{2} \Phi^2 + \sum_{n=1}^\infty \frac{a_n}{n} \Phi^n +  \frac{1}{2} \sum_{n=1}^\infty c_n \Phi \Box^n_g \Phi + ...  \right),
\end{equation}
with additional terms proportional to the Ricci curvature of $g$, possible higher derivative terms and terms arising from the conformal anomaly \cite{de Haro:2000xn,Skenderis:2002wp,Bianchi:2001kw,Skenderis:2008dg}. In the standard quantisation, $\Delta_+ = d-\Delta_-$ is the CFT operator dimension and $\Box^n_g$ is the d'Alembertian operator on the metric $g_{\mu\nu}$. As usual, we use $\Delta_\pm = \frac{d}{2} \pm \nu = \frac{d}{2} \pm \sqrt{ \left(  \frac{d}{2} \right)^2 + m^2 }$. Note that we will be working in standard Dirichlet quantisation, which will be reflected in the identification of $\phi$ as the source of the dual operator $\CO$, of which the vacuum expectation $\langle \CO \rangle$ is determined by $\Pi$. At $\rho_0$, the source is related to the bulk scalar  as $\Phi =  \rho_0^{\Delta_-} \phi$. Despite the fact that the structure of multi-trace operators \cite{Witten:2001ua,Akhmedov:2002gq,Mueck:2002gm,Minces:2002wp} in the effective action is more apparent in mixed and alternative (Neumann) quantisations, we wish to avoid limitations imposed by unitarity on the interval of operator dimensions where only $\Delta_\CO = \frac{d}{2} - \nu$ with $\nu\in [0,1]$ are allowed. Mixed quantisation is a hybrid of Dirichlet and Neumann boundary conditions, but behaves as alternative at the initial cut-off. The vacuum of the theory runs towards the Dirichlet quantisation \cite{Vecchi:2010dd,Hartman:2006dy}. Dirichlet and Neumann quantisations are related by a Legendre transformation and their connection has been explored in many references, among them \cite{Faulkner:2010jy,Klebanov:1999tb,Papadimitriou:2007sj,Gubser:2002vv}. 

Under the RG transformation $\rho \to \rho + \delta \rho$, we begin with a flow from $\rho_0$. The bulk action will change as we reduce the bulk by integrating out slices of geometry. Since we are working with large-$N$ theories, this in practice means that $S_\text{bulk}$ only changes its integration region $[\rho, \infty)$ to $[\rho + \delta\rho, \infty)$ in \eqref{FullBareBulkAction}. In the Wilsonian RG, the partition function of the bare action is kept fixed under the flow. Analogously, we will fix the full action $S$ from \eqref{FullBareBulkAction} so that $S_\text{B}$ will flow to compensate for the change of $S_\text{bulk}$. We saw that $S_\text{B} [\rho_0]$ was the same as the semi-classical path integral contribution from $r \in [0,\rho_0]$, implying that $S$ defines the entire bulk theory as well as its dual. Dual scalar operators $\CO$ will therefore run as bare operators in quantum field theory, and with a sensible definition of the wavefunction renormalisation, they will keep the renormalised operators invariant. Using the relation between the bulk scalar $\Phi$ and the source $\phi$, we can generalise their relation to permit for the running of the bare source. We define
\begin{equation}\label{PhiZDef}
\Phi (\rho) = \rho^{\Delta_-} Z (\rho) \phi(\rho),
\end{equation}
with $Z(\rho_0) = 1$. On dimensional grounds, from definition \eqref{BarePi}, we expect $\Pi_\text{B}$ to transform under the scaling RG transformation, $\rho \to \rho = \rho_0 + \delta\rho$, as
\begin{equation}\label{PiBZTransform}
\Pi_{\text{B}} (\rho) = \left( \frac{\rho}{\rho_0} \right)^{-\Delta_+} Z(\rho) \Pi_{\text{B}} (\rho_0).
\end{equation}

Finally, we generalise the bare boundary action constructed in \eqref{SubBdyAct} to permit for all its coefficients and operators to run along the flow. We also, at this point, take the $\rho_0 \to 0$ limit. The action becomes
\begin{equation}\label{BareActRho}
S_\text{B} [\rho] = \alpha (\rho) + \!\! \int \limits_{r=\rho} \!\!  d^d x \sqrt{-g} \left[ \frac{1}{2} \frac{\Pi}{\sqrt{-g}} \Phi - \sum_{n=2}^{\infty} \frac{1}{n} \lambda_n \Phi^n  \right],
\end{equation}
where $\alpha$, $\Pi$ and $\lambda$ are now functions of $\rho$, which will run under the RG flow equations. In a fixed background $G_{MN}$, the induced $g_{\mu\nu}$'s on $d$-dimensional boundaries are fixed functions of $r$, and hence $\rho$. Polynomial terms directly correspond to counter-terms permitting for a potentially necessary series of such terms. Their structure is completely determined by holographic renormalisation and each term transforms under the bulk isometries. This is crucial, as bulk isometries correspond to symmetries of the dual gauge theory and we expect only such terms to arise in the Wilsonian effective action. $S_{\text{B}} [\rho]$ can then be considered as the effective action of the Wilsonian renormalisation group of composite operators $\CO$. In alternative quantisation where $\Phi \sim \CO$, the multi-trace structure becomes immediately apparent and each $\Phi^n$ term corresponds to the $\CO^n$ effective term. Furthermore, the holographic counter-terms determine the initial values of the effective cosmological constant $\alpha$, the running conjugate momentum $\Pi$ and coefficients $\lambda_n$ at the start of the flow:
\begin{align}\label{CoeffBC}
\alpha (0) = 0,&	&\Pi(0) = \Pi_0,& &\lambda_2(0) = \Delta_- = a_2,& & \lambda_n (0) = a_n.
\end{align}
Notice that we did not include any derivative or logarithmic terms in \eqref{BareActRho}. Compared to the original counter-terms \eqref{CTBdyAct}, this vast simplification in the structure of the effective terms is possible because of a well-defined $\rho_0 \to 0$ limit resulting from the asymptotically AdS structure of considered spacetimes. For well-behaved bulk scalars and their derivatives near the initial arbitrary $\rho_0$, as ensured by the boundary condition and a smooth metric, derivative terms $\Phi^l \Box^n_g \Phi^m$ vanish in momentum space. To see this, consider first the $c_1 \Phi \Box_g \Phi$ counter-term. In the momentum space representation of $\Phi$, $e^{i k\cdot x}$ involves a contraction with the Minkowski metric and not the induced $g_{\mu\nu}$. The counter-term $c_1 \Phi \Box_g \Phi |_{r=\rho_0}$ therefore gives the $c_1 \rho_0^2 k^2 \Phi(k) \Phi(-k)$ momentum space contribution to the effective action in asymptotically AdS spaces. On the other hand, the coefficient in the $\Delta_{-} \Phi(k) \Phi(-k)$ term includes no factors of $\rho_0^2$. Hence, the $c_1 \Phi \Box_g \Phi$ counter-term vanishes in the limit of $\rho_0 \to 0$, where we wish to begin the RG flow at the extreme UV fixed point. We see that the boundary condition \eqref{CoeffBC} on $\lambda_2$ would still apply, even in the explicit presence of an added double-trace derivative term. The running of $c_1$ can thus be simply absorbed in $\lambda_2$. For the same reason, other counter-terms with derivatives also vanish in the limit. The logarithmic terms coming from the conformal anomaly vanish in this limit as well because they typically appear inside the coefficients of derivative terms. The coefficients of such terms behave as $\rho^n_0 ( \ln \rho_0)^m \to 0$ when $n\geq 1$ and so we may again absorb them into $\lambda_n$'s without affecting the RG flow. Another reason for this simplification is that in the RG flow equations, only derivatives with respect to $\rho$ appear and not the momentum. Additional momentum terms in the effective action $S_\text{B}$ therefore do not matter in this setup and can in general be thought of as absorbed into $\lambda_n$'s. We conclude that the polynomial structure of $\Phi^n$ terms in \eqref{BareActRho} is sufficient to account for the entire RG flow. 

An example where this structure fails is the Coulomb branch of the $\CN=4$ theory, which has a counter-term $\left(1 + 1/\ln \rho_0^2 \right)\Phi^2$ at the initial $\rho_0$ cut-off boundary \cite{Bianchi:2001kw}. Further care then needs to be taken when specifying the boundary conditions for $\lambda_n$ and we expect that in some cases the divergent counter-terms prevent us from taking $\rho_0 \to 0$ altogether. The flow must then begin at $\rho_0 > 0$. This is, of course, not surprising in theories with a Landau pole, which may arise particularly in some scenarios on the Coulomb branch. We will not consider such examples in this paper and in our case the above simplification applies to all studied RG flows. 

The running coefficients $\lambda_n$ will become functions of $\rho^2 k^2$ as well as other physical scales along the flow. They can therefore be expanded in powers of the momentum. This implies that when we transform them back to position space, the bare effective action \eqref{BareActRho} will organise itself as the gradient expansion with cut-off dependent terms $\frac{d_1}{\Lambda^2} \CO  \Box \CO + ... +  \frac{ d_n}{\Lambda^{2n}} \CO \Box^n \CO + ...  $. We will therefore still find an infinite series of derivative terms, as is expected in the Wilsonian effective action, despite our only keeping the non-derivative terms in \eqref{BareActRho} as $\rho_0 \to 0$. Momentum dependence will arise from the derivative kinetic term in the bulk action and descend to all of the $\lambda_n(k)$'s through the coupled differential equations for $\lambda_n$'s. Note that we are only working with scalar bulk theories with two-derivative Lagrangians. In theories with higher derivatives, which may arise from supergravity, additional complications would occur.

\subsection{Renormalisation group equations}\label{Sec:RG2}

With an effective boundary action $S_\text{B}[\rho]$ which we built from the structure of holographic renormalisation, we now follow \cite{Faulkner:2010jy} and derive renormalisation group equations for the flow of \eqref{BareActRho}. They are obtained by varying the position of the brane, $\rho \to \rho + \delta\rho$, and insisting that the overall bare action \eqref{FullBareBulkAction} remains constant at any $r=\rho$:
\begin{equation}\label{HFlowEq}
\partial_\rho S_\text{B} [\rho] = - \!\! \int \limits_{r=\rho} \!\! d^d x \; \CH = -  \! \! \int \limits_{r=\rho} \!\! d^d x \left( \frac{\delta S_\text{B}}{\delta \Phi} \frac{\partial \Phi}{\partial r} - \sqrt{-G} \CL \left(\Phi, \partial_M \Phi \right) \right).
\end{equation}
Since we are neglecting the metric backreaction, components of $G_{MN}$ are simply treated as functions of $r$. $\CH$ stands for the Hamiltonian density with $r$ treated as time, making this a Schr\"{o}dinger-type evolution equation \cite{Mansfield:1999kk,Heemskerk:2010hk}, or a Hamilton-Jacobi equation \cite{de Boer:1999xf,Papadimitriou:2004ap}. Using definition \eqref{BarePi}, equation \eqref{HFlowEq} can be rewritten as
\begin{equation}\label{HFlowEq2}
\sqrt{G^{rr}(\rho)} \partial_\rho S_\text{B} =  - \!\! \int \limits_{r=\rho} \!\! d^d x \sqrt{ -g} \left( \frac{1}{2 g} \left( \frac{\delta S_\text{B}}{\delta \Phi} \right)^2   +  \frac{1}{2} g^{\mu\nu} \partial_\mu \Phi \partial_\nu \Phi + V(\Phi) \right),
\end{equation}
where in case of a singularity in $G_{MN}$, we must be more careful before dividing both sides by $G^{rr}$. Now since we are working with Dirichlet boundary conditions, we can insist on the bulk scalar $\Phi$ to remain constant on the moving brane surface throughout the flow, i.e. $\partial_\rho \Phi = 0$. This is equivalent to a general Dirichlet problem \cite{Brattan:2011my} and related to the more general analysis of \cite{Kuperstein:2011fn}. Despite fixing $\Phi$, $S_\text{B}$ still has enough structure for the flow to run. Furthermore, it is important to keep in mind that the source $\phi$, as defined in \eqref{PhiZDef}, will still run. Matching the powers of $\Phi$, we derive the following set of differential equations describing the RG flow: 
\begin{equation}\label{PiRGeq}
\sqrt{G^{rr} }  \partial_\rho \Pi =  - 2 \lambda_2 \Pi  ,
\end{equation}
\begin{align}
\frac{1}{\sqrt{-G}} \partial_\rho \left( \sqrt{-g} \lambda_2 \right)    	=&      	- \lambda_2^2  + k_\mu k^\mu  +m^2 + \frac{2}{\sqrt{-g}} \lambda_3 \Pi,        \label{Coeff2RGeq}   \\
\frac{1}{\sqrt{-G}}  \partial_\rho \left( \sqrt{-g} \lambda_n \right)		=&		- \frac{n}{2}   \sum_{m=2}^{n}  \lambda_m \lambda_{n+2-m}  + \frac{n}{\sqrt{-g}} \lambda_{n+1} \Pi   + b_n	\label{CoeffnRGeq}
\end{align}
and
\begin{equation}\label{AlphaRGeq}
\frac{1}{\sqrt{-G}} \partial_\rho \alpha  = - \frac{1}{2 g} \int \frac{d^d k}{(2\pi)^d}  \Pi(k) \Pi(-k).
\end{equation}
The factor of $2$ in \eqref{PiRGeq} comes from the $\delta\Pi / \delta\Phi$ variation between two canonical conjugates. Given that we fixed $\Phi$, it is not surprising that $\Pi$ has to incorporate the total running of the action, proportional to two units of $\Phi$. $\Pi$ therefore has to run differently than the bare $\Pi_{\text{B}}$ in \eqref{PiBZTransform}. We could have equally well used the flux factor term $\CF \Phi^2$, to which we come in section \ref{Sec:RG3}, instead of $\Pi$. If we had treated $\Phi$ and $\Pi$ as independent, then the differential equation \eqref{PiRGeq}, without the factor of $2$, would hold for $\Pi$. The full \eqref{PiRGeq} would describe the running of $\Pi^2$. The analysis of these equations in various fixed backgrounds will be the subject of the following sections.

\subsection{Two-point correlation functions and the Callan-Symanzik Equation}\label{Sec:RG3}

To connect the RG equations \eqref{PiRGeq}-\eqref{AlphaRGeq} coming from the bulk to boundary physics, we use the fact that a renormalised two-point correlation function is completely determined by the canonical conjugate momentum $\Pi$ \cite{Papadimitriou:2004ap,Papadimitriou:2004rz,Iqbal:2008by}. It is given by a flux factor $\CF$,
\begin{equation}\label{GAsFlux}
G (k) \equiv \left\langle \CO(k) \CO(-k) \right\rangle = \CF (\rho,k), ~ \text{where} ~ \CF = \Pi / \Phi.
\end{equation}
The type of the two-point function (retarded, advanced, etc.) is determined by boundary conditions imposed on $\Phi$ and consequently $\Pi$. $\CF$ is extracted from the renormalised boundary action by taking functional derivatives with respect to the source $\phi$ \cite{Gubser:1998bc,Witten:1998qj,Son:2002sd,Herzog:2002pc},
\begin{equation}\label{GFromDerivatives}
G(k) =\rho^{2 \Delta_-} Z^2 (\rho)  \frac{\delta}{\delta\Phi} \frac{\delta}{\delta\Phi} \int \!\! d^d x \; \Phi \CF(\rho) \Phi,
\end{equation}
having used equation \eqref{PhiZDef}. $G(k)$ is the fully renormalised, scale-independent two-point function. Therefore $dG / d\rho= 0$. Since everything in our equations is explicitly written out in $\rho$, we can interchangeably use $d/d\rho$ and $\partial_\rho$. Acting with $\rho\partial_\rho$ on \eqref{GFromDerivatives} we have
\begin{equation}\label{DOnG}
\rho \partial_\rho \left( \frac{  \rho^{2\Delta_-} Z^2 (\rho)   \Pi(\rho)}{\Phi} \right) = 0.
\end{equation}
It then follows from initial conditions \eqref{CoeffBC} that $\lambda_2$ can be conveniently written as
\begin{equation}\label{L2Def}
\lambda_2 (\rho) = \Delta_- + \gamma(\rho), ~ \text{with}~ \gamma(0) = 0.
\end{equation}
Using $\partial_\rho \Phi = 0$ along with the RG equation \eqref{PiRGeq} for $\Pi$ thus gives
\begin{equation}\label{GammaAsZ}
\gamma =  \frac{ \partial \ln Z}{\partial \ln \rho}
\end{equation}
for metrics, such as pure AdS in \eqref{AdSMet1}, with $G^{rr} = r^2$. Equation \eqref{GammaAsZ} reveals the expected relation between the anomalous dimension $\gamma$ of the dual field theory operator $\CO$ and the wavefunction renormalisation of $\Phi$, and consequently $\Pi$. A cut-off dependent two-point function, disposing of the $\rho^{2\Delta_-}$ factor which is a consequence of the conformal (scale-invariant) scaling, obeys the Callan-Symanzik renormalisation group equation
\begin{equation}\label{CSEq1}
\left( \rho \frac{\partial}{\partial \rho} + 2 \gamma \right) \tilde G (\rho,k) = 0.
\end{equation}
Two-point function $\tilde G$, which is proportional to $\Pi$, had been holographically renormalised in our construction and is therefore finite in the $\rho_0 \to 0$ limit. Its solution has the standard form of a running two-point function,
\begin{equation}
\tilde G (\rho, k) = \lim_{\rho_0\to 0} \tilde G (\rho_0, k) \exp \left\{ - 2 \int_{r=\rho_0}^{r=\rho} \gamma(r) d \ln r \right\},
\end{equation}
so that $\tilde G(\rho) = Z^{-2} (\rho) \tilde G (\rho = 0)$. The full running dimension of the operator $\CO$ is therefore $\Delta_\CO = \Delta_+ - \gamma$.

Setting $\lambda_n = 0 $ for $n \geq 3$, we see that the equation \eqref{Coeff2RGeq} becomes the renormalisation group equation for the anomalous operator dimension. Alternatively, this equation also describes the flow of the double-trace coupling $f$ in $f \CO\CO$, with $\rho \partial_\rho \gamma$ proportional to its beta function $\beta_f$ \cite{Vecchi:2010dd,Faulkner:2010jy,Heemskerk:2010hk}. Given that \eqref{Coeff2RGeq} is a quadratic equation, we anticipate $\beta_f \propto \gamma^2$. These results are indeed consistent with field theory calculations of large-$N$ theories in \cite{Pomoni:2008de}, where it was found that $\gamma \propto f$ and $\beta_f \propto f \gamma \propto \gamma^2$ in large-$N$ theories conformal in their single-trace sector. Furthermore, we can now justify setting $\lambda_{n\geq3} = 0$. Coefficients $\lambda_n$ are similarly related to multi-trace couplings of the $n$-th order. In large-$N$ analysis, we may safely neglect all deformations of order higher than two (double-trace), as all such deformations are sub-dominant in the large-$N$ limit \cite{Pomoni:2008de}. Another reason is that by working in the standard (Dirichlet) quantisation, the smallest possible operator dimension is $\Delta_\CO = d/2$, implying that any triple-or higher-trace operator would necessarily be irrelevant. Even in the alternative (Neumann) quantisation, at the unitarity bound with $\Delta_\CO = d/2-1$, such an operator could at most have a marginal triple-trace deformation in $d=6$ dimensions.
 
In scenarios when $G^{rr}$ is a more complicated function of $r$, we need to redefine the quantities considered above. We write
\begin{equation}\label{L2DefGrr}
\lambda_2 (\rho) = \frac{\sqrt{ G^{rr} (\rho) }}{\rho} \left( \Delta_- + \gamma(\rho) \right).
\end{equation}
To demonstrate that this is a sensible definition, we consider an example of a near-extremal D3 brane generalised to any dimension, which gives temperature to the dual of the pure AdS geometry. This spacetime is also known as the black brane in AdS and has the metric of form
\begin{equation}\label{BBAdSMetric0}
ds^2 = \frac{L^2}{r^2} \left( - f(r) dt^2 + d\vec{x}^2_{d-1} + \frac{dr^2}{f(r)} \right),
\end{equation}
with the thermal factor behaving as $f(r) \to 1$ when $r \to 0$. We will set the AdS radius $L$ to $L=1$. The metric \eqref{BBAdSMetric0} is general enough to describe all cases with non-zero temperatures, which we will analyse in section \ref{Sec:AnomDim}. With $G^{rr} = r^2 f(r)$ in \eqref{BBAdSMetric0}, the expression \eqref{L2DefGrr} takes the form 
\begin{equation}\label{L2DefT}
\lambda_2 (\rho) = \sqrt{ f (\rho) } \left( \Delta_- + \gamma(\rho) \right),
\end{equation}
which still obeys the initial condition \eqref{CoeffBC}. Equation \eqref{PiRGeq} then loses its explicit dependence on the thermal factor, giving as before,
\begin{equation}
\left( \rho \frac{\partial}{\partial \rho} + 2 \Delta_- + 2 \gamma \right) \Pi = 0.
\end{equation}
Using \eqref{L2DefT} in \eqref{Coeff2RGeq} will, however, modify the flow equation for the anomalous dimension $\gamma$ by adding thermal corrections to it. Relation \eqref{L2DefT} with its thermal rescaling can also be understood directly from the equation of motion, $\frac{1}{\sqrt{-G}} \partial_r \left(\sqrt{-G} G^{rr} \partial_r \Phi \right) + G^{\mu\nu} \partial_\mu \partial_\nu \Phi - m^2 \Phi = 0$, for a massive scalar field. We follow the standard procedure for determining the operator dimensions in AdS/CFT by analysing solutions near the asymptotically AdS boundary ($r \ll 1$). In pure AdS with the metric \eqref{AdSMet1}, $\Phi$ there behaves as $r^\Delta$. Using the equation of motion then gives a relation $\Delta (\Delta - d) = m^2$, where in our Dirichlet (standard) quantisation we take $d - \Delta \equiv \Delta_+$ to be the dimension of the dual operator $\CO$ to $\Phi$. The other, smaller, solution is our $\Delta_- \equiv \Delta$. Using the same behaviour of $\Phi$ in the black brane background \eqref{BBAdSMetric0} near $r \approx 0$, which is justified by the asymptotic structure and the Fefferman-Graham expansion \cite{Fefferman:1985}, immediately yields a modified relation $\Delta (\Delta - d) f(r) = m^2$. Now of course in the limit of $r \to 0$, $f(r) \to 1$, and so the dimension of the dual operator at the AdS infinity stays independent of the thermal factor. But this identity suggests that the thermal factor will contribute to the dimensions of scalar operators as we flow into the bulk, towards the horizon. It gives a thermal modification of $\Delta_\pm \to \sqrt{f(r)} \Delta_\pm$, with the same scaling as in \eqref{L2DefT}. The radial derivative in the bulk is related to the dilatation operator on the field theory side of the duality \cite{Papadimitriou:2004ap,Papadimitriou:2004rz}. Hence, we can think of the rescaled anomalous dimension as arising from the red-shifted dilatation operator. Energies in the field theory and thus the operator scaling dimensions are measured as being conjugate to the proper time on the flowing hypersurface at $\rho$.\footnote{We thank the JHEP referee for pointing this out.}

\section{Anomalous dimensions and double-trace deformations}\label{Sec:AnomDim}

In this section, we study the renormalisation group equation \eqref{Coeff2RGeq}, describing the radial evolution of the anomalous operator dimension induced by effective double-trace deformations. We aim to formulate a method for determining a precise functional dependence, $\Lambda(\rho,...)$, between energy scales in the brane's QFT and their dual bulk quantities. Neglecting all but double-trace deformations, as argued in section \ref{Sec:RG3}, the relevant renormalisation group equation is given by  
\begin{equation}\label{DTRGEq}
\frac{1}{\sqrt{-G}} \partial_\rho \left( \sqrt{-g} \lambda_2 \right)    = 	- \lambda_2^2   - \left( \frac{d}{2} - \nu \right) \left(\frac{d}{2} + \nu \right) + g^{\mu\nu} \eta_{\mu\rho} \eta_{\nu\sigma} k^\rho k^\sigma,
\end{equation}
which we will analyse in various asymptotically AdS backgrounds. We have explicitly written out contractions of the physical brane momentum $k^\mu$ with the flat Minkowski metic $\eta_{\mu\nu}$. We use the previously established definitions of $\lambda_2$, i.e. $\lambda_2 (\rho) = \Delta_- + \gamma(\rho)$, and $\lambda_2(\rho) = \sqrt{f(\rho)} \left( \Delta_- + \gamma(\rho) \right)$ in thermal cases. 

For a well-defined and realistic RG flow, it is essential that we introduce physical conditions on the behaviour of $\lambda_2$. Understanding how $\lambda_2$ should behave at two ends of the flow will then translate into the necessary boundary conditions required to extract the dependence of the cut-off on the bulk from \eqref{DTRGEq}. 

In quantum field theory, the wavefunction renormalisation $\CZ$ is interpreted as the probability of finding a bare particle in a physical one-particle state \cite{Higashijima:2003et,Peskin:1995ev}. This can easily be understood from the K\"{a}ll\'{e}n-Lehmann spectral representation of a two-point function. As a consequence, $\CZ$ must satisfy the unitarity bound $0\leq \CZ \leq 1$, or equivalently, the spectral density function has to remain positive. Perturbatively, $\CZ = 1 -  g^2 A \ln \Lambda^2 / \mu^2 $ up to one-loop, where $\mu$ is the renormalisation scale and the ratio of $\Lambda/\mu$ is always required by dimensional analysis. In theories that are conformal in their single-trace sector, all higher-loop corrections are subleading in the large-$N$ expansion \cite{Witten:2001ua,Pomoni:2008de}. Therefore, $\CZ = 1$ at $\Lambda = \mu$. The renormalisation scale $\mu$ can in our construction be identified with the physical Lorentzian momentum $\sqrt{-k^2}$ at which we are probing the operator. This is because we have used the same "minimal subtraction" scheme in holographic renormalisation at the initial cut-off $\Lambda_0 (\rho_0)$ for any value of the physical operator momentum of interest. Equivalently to the discussion of the wavefunction renormalisation, a bare two-point function of an operator with dimension $\Delta$ can be schematically expanded around its anomalous dimension, to the leading order in large-$N$, as
\begin{equation}\label{PertAnomDimExp}
\left\langle \CO(x) \CO(0) \right\rangle_\Lambda = \CZ^2 \left\langle \CO(x) \CO(0) \right\rangle = \frac{c}{x^{2 \Delta}} \left(1 - 2 g^2 A \ln \Lambda^2 x^2 + ... \right).
\end{equation}
Since $x$ scales as $1/\sqrt{-k^2}$, we again recover that $\CZ = 1$ at $\Lambda = \sqrt{-k^2}$. In general, we expect both expansions for $\CZ$ and $\left\langle \CO(x) \CO(0) \right\rangle_\Lambda$ to have all higher-order terms proportional to powers of $\ln \Lambda^2 / k^2$. Hence, we expect that $\CZ$ always takes the value of $\CZ = 1$ at $\Lambda = \sqrt{-k^2}$, to all orders in large-$N$. 

More generally, the theories we wish to consider all have conformal fixed points in the extreme UV. These large-$N$ theories, initially conformal in their single-trace sector, will only be perturbed by relevant IR mass and thermal scale deformations. Furthermore, since we are working with the standard (Dirichlet) quantisation, double-trace deformations in the effective Wilsonian action will be irrelevant and therefore only affect the UV regime. Now, quantities such as the anomalous dimension must become cut-off independent at the end of the RG procedure. This is ensured by imposing the appropriate renormalisation condition, which fixes the value of $\Delta_\CO$ with respect to the running cut-off. We therefore demand that $d \gamma / d \Lambda = 0$ at the RG scale $\mu$. It is natural, as discussed above, that this scale should equal the momentum scale $\sqrt{-k^2}$ of the operator $\CO$. The $\sqrt{-k^2} = \mu$ scale must thus be the minimal scale down to which we can integrate out higher momentum modes, even in the presence of IR scales below the operator momentum. In extracting the cut-off dependence on the bulk, we will not consider any examples with physical momenta below the IR mass scales, which would require us to integrate out those scales in order for the cut-off to run down to $\sqrt{-k^2}$. In all cases we wish to consider, the running Wilsonian cut-off scale $\Lambda$ can only exist in the interval of $\Lambda \in [\Lambda_{\times} , \Lambda_0 \to \infty ]$, where the lowest possible cut-off $\Lambda_\times$ satisfies $\Lambda_\times = \mu = \sqrt{-k^2}$. 

Given our holographic construction of the Wilsonian renormalisation group in section \ref{Sec:RG}, the cut-off should become a function of the running scale $\rho$. It thus follows that in order for the physical operator dimension to be $\rho$-independent, we must impose the renormalisation condition
\begin{equation}\label{RGCon1}
\frac{\partial \gamma}{\partial \rho}  = 0 ~\text{at}~\Lambda_\times(\rhox) = \mu = \sqrt{-k^2},
\end{equation}
while keeping all other scales fixed. The lowest possible value of the Wilsonian cut-off $\Lambda_\times$, given some physical timelike operator momentum $\sqrt{-k^2}$, must therefore be a function of the largest possible value of $\rho$ that the brane can reach while flowing into the bulk. The variable $\rhox$ is used throughout to indicate the value of $\rho \in [\rho_0\to 0,\rhox]$ where the RG flow must terminate.

When $\rho$ and $k$ are the only two scales present in the theory, as in the case of the conformal $\CN=4$ theory, then each of the dimensionless running quantities $Z$ and $\gamma$ can only depend on the dimensionless product $\rho^2 k^2$. Therefore $\rho  \partial \gamma/\partial \rho= \sqrt{-k^2}  \partial \gamma / \partial \sqrt{-k^2}$, which implies that the renormalisation condition \eqref{RGCon1} can be rewritten as $\partial\gamma / \partial \sqrt{-k^2} = 0$ when $\rho \neq 0$ and $\sqrt{-k^2} \neq 0$. In such a theory, the physics at the cut-off, where $\sqrt{-k^2} = \Lambda_\times$, remains conformal and only becomes modified by double-trace deformations below the cut-off scale, i.e. $\sqrt{-k^2} < \Lambda_\times$. This follows from the fact that in large-$N$ theories conformal in their single-trace sector the beta function of the double-trace coupling $f \CO^2$ behaves as $\beta_f  \propto \gamma^2 \propto \mu \partial_\mu \gamma$ \cite{Pomoni:2008de}. If such a theory is deformed by IR energy scales such as a mass scale or temperature, the renormalisation group condition \eqref{RGCon1} must still hold. The theory will still possess a conformal fixed point in the extreme UV, but the physics at the lowered cut-off scale will no longer be conformal. To see this, let us introduce a mass scale $\CM$, which is dual to a bulk scale at some radial position $r_1$, such that $\CM \propto 1/r_1$. Function $\gamma$ can now depend on three dimensionless combinations $\rho^2 k^2$, $r_1^2 k^2$ and $\rho^2 / r_1^2$, which implies that  $\rho  \partial \gamma/\partial \rho= \sqrt{-k^2}  \partial \gamma / \partial \sqrt{-k^2} - r_1 \partial \gamma / \partial r_1$. The RG condition \eqref{RGCon1} then fixes $ \sqrt{-k^2} \partial \gamma / \partial \sqrt{-k^2} = r_1 \partial \gamma / \partial r_1 \neq 0$ at $\Lambda_\times = \sqrt{-k^2}$, which shows that the theory is no longer scale-invariant at the running cut-off. It is easy to see that the same argument can be extended to the presence of several mass scales and temperature. 

Let us now address the question of what value the anomalous dimension $\gamma(\rhox)$ should take when we impose the renormalisation conditions. The second renormalisation condition must reflect the fact that the bare two-point function, which depends on $\Lambda_0 (\rho_0)$, should be $\Lambda(\rho)$-independent at the renormalisation scale $\mu = \sqrt{-k^2}$. This can be justified by the fact that the only divergences present in theories under consideration are the UV divergences, which are removed by the minimal subtraction at $\rho_0$. No additional IR divergences affect the bare two-point function, nor the anomalous dimension. Since we acquire no new divergences in the process of integrating out the slices of geometry, we can therefore set the bare two-point function to its initial value at the scale $\mu$ where we impose the two renormalisation conditions. In our holographic construction, the bare running two-point function must thus be $\rhox$-independent at the scale $\Lambda_\times= \mu = \sqrt{-k^2}$ where the flow terminates. With these observations in mind we can find $\gamma(\rhox)$ using the results from sections \ref{Sec:RG2} and \ref{Sec:RG3}. The bare operator $\CO$ scales proportionally to $\Pi_\text{B}$ and not $\Pi \propto \CF \propto \tilde G$, which runs as the holographically renormalised, scale-dependent two-point function. Namely,
\begin{equation}\label{OwithZ}
\CO (\rho) = \left( \frac{\rho}{\rho_0} \right)^{-\Delta_+} Z(\rho) \CO(\rho_0).
\end{equation}
$Z(\rho)$ in our construction is analogous to $\CZ$ in a general field theory, but without the LSZ normalisation. It quantifies the renormalisation group - scale transformations. The generating functional of the dual field theory $\left\langle \exp \{ \int \!\! \CO \phi \} \right\rangle$ gives $\langle \CO \CO \rangle$ after taking two functional derivatives with respect to $\phi$. Using \eqref{PhiZDef} with $Z(\rho_0)=1$ and $\partial_\rho \Phi = 0$, as well as \eqref{OwithZ}, we see that the two-point function on the field theory side scales as
\begin{equation}\label{OOwithZ}
\left \langle \CO(k) \CO (-k) \right\rangle_\rho \propto \frac{\delta}{\delta \phi} \frac{\delta}{\delta\phi} \int \!\! d^d x \; \CO \phi = \left(\frac{\rho}{\rho_0} \right)^{-2 \nu} Z^2 (\rho) \frac{\delta}{\delta \phi_0} \frac{\delta}{\delta\phi_0} \int \!\! d^d x \; \CO_0 \phi_0.
\end{equation}
It therefore follows that
\begin{equation}\label{CutoffCond1}
Z (\rhox) = \left(\frac{\rhox}{\rho_0} \right)^{\nu},
\end{equation}
where the flow terminates at $\rhox$. The bare two-point function $\left \langle \CO(k) \CO (-k) \right\rangle_\rho$ remains divergent when we take $\rho_0 \to 0$. This divergence is, as discussed above, completely removed by the minimal subtraction through the holographic renormalisation of $\CO$ and $\phi$. We can therefore see that the second renormalisation condition is completely analogous to perturbatively finding that $\CZ = 1$  in expression \eqref{PertAnomDimExp}, even in the presence of additional relevant deformations. Their presence will only be reflected in the functional dependence of $Z$. Using \eqref{L2Def} and \eqref{GammaAsZ}, the boundary conditions for $\gamma$ at the two ends of the flow become
\begin{align}\label{GammaBC}
\gamma(0) = 0&	&\text{and}&	&\gamma_\times \equiv \gamma(\rhox) = \nu.
\end{align}
It is worth noting that the rescaling in our construction works in the opposite way compared to the usual Wilsonian RG. Usually one rescales the theory upwards after integrating out high momentum modes above $\Lambda$. The rescaled effective theory thus remains defined up to the initial $\Lambda_0$ and the anomalous dimension can be extracted from the rescaling of the field variable. We are, on the other hand, rescaling the original ($\sqrt{-k^2} \leq \Lambda_0$) fields and operators down to the effective theory defined up to $\Lambda$.

It follows from \eqref{GammaBC} that the effective operator dimension at $\rhox$ is $\Delta_\CO = d/2$. This is true regardless of our choice of quantisation. The coupling $f$ in the effective double-trace term $f \CO^2$ will consequently have its mass dimension equal to zero at $\rhox$. In the absence of mass scales or temperature, it is easy to see that this fact is consistent with our expectation that an effective field theory of a single-trace UV conformal theory should remain conformal at the maximal allowed momentum, right at the cut-off scale. This is because the conformal anomaly (trace anomaly) arises when UV divergences make coupling constants acquire non-zero anomalous mass dimensions \cite{Callan:1970ze}. The breaking of conformal invariance by quantum fluctuations is therefore a purely UV effect. For operator momenta at the cut-off scale, however, all UV modes had been completely removed, which ensures a zero coupling dimension at all loops. Below the cut-off, when $\sqrt{-k^2} < \Lambda_\times$, operators obtain non-trivial anomalous dimensions depending on their momentum. The double-trace coupling also acquires a non-zero mass dimension. In the presence of relevant mass scales a similar argument can be repeated for the behaviour at the cut-off if we consider rescaling the entire theory along a symmetry of $\gamma$ and $Z$, including all the IR scales, up to the initial extreme UV fixed point at $\Lambda_0$. This works as long as we stay above the IR scales and as long as we have not introduced any new IR divergences. The value of $\gamma(\rhox)$ should stay the same in those cases even though the theory will no longer be at a fixed point for $\Lambda < \Lambda_0$, where now $\partial \gamma_\times / \partial \sqrt{-k^2} \neq 0$.  

For a well-defined RG flow, having already specified the renormalisation conditions, we can also impose the condition that $\gamma$ must remain real, as flowing to a complex operator dimension would imply an unstable theory. Such dynamical symmetry breaking would also inevitably break conformal symmetry \cite{Pomoni:2008de}. Furthermore, $\gamma$ must not break the unitarity bound $-\infty < \gamma \leq \nu + 1$, which follows from $d/2 - 1 \leq \Delta_\CO = d/2 + \nu - \gamma$.

We will see from the analysis that $\gamma$ never decreases throughout the flow, so that $\partial_\rho \gamma \geq 0$. This is expected as it runs between $\gamma = 0$ and $\gamma = \nu$, ending with $\partial_\rho \gamma_\times = 0$. Furthermore, in our examples, there is no reason to expect the theory to run into another fixed point along the flow, so that $\partial_\rho \gamma > 0$ for $\rho \in (0,\rhox)$. This immediately implies that $\gamma \geq 0$ for all $\rho$. 

The only exception to this behaviour are the runnings of lightlike (or vacuum) operators which we will have to treat separately. As is usual in quantum field theory, there is no smooth limit of momentum $k^2 \to 0$ that we could take to simply recover the physics of $\CO(k^2=0)$ from timelike cases. Anyhow, since lightlike operators should be able to run into extreme IR of the bulk ($\rho\to\infty$), there is no reason to expect that our anaysis, when using sharp IR cut-offs which abruptly terminate the flow, would apply to such lightlike runnings and set $\gamma = \nu$ and $\partial_\rho \gamma = 0$ at the point where the flow terminates at the horizon or a general metric singularity. This applies to all cases, the GPPZ, the black brane, etc., with the exception of the undeformed pure AdS scenario, where we can flow until $r\to\infty$ without running into singularities. In that case, it is natural to expect that $\partial_\rho \gamma (\rho\to\infty) = 0$, so that the flow ends at an IR fixed point. As for the relevance of vacuum states, we lastly note that we will always subtract off the $k^2 = 0$, $\langle \CO(0) \CO(0) \rangle$ vacuum contribution to correlators $\langle \CO(k) \CO(-k) \rangle$, which means that $\gamma (\rho,k) = 0$ at $k^2 \to 0$. Furthermore, since the effective double-trace deformations are irrelevant, the anomalous dimension at low momenta will not be affected by them.

With these observations in mind, we impose the following four conditions that the behaviour of $\gamma$ is expected to follow and from which we can extract the dependence of $\Lambda$ on the bulk by identifying where the flow terminates:

\textit{i)}	$\frac{\partial \gamma}{\partial \rho} = 0$ and $\gamma = \nu$ at $\sqrt{-k^2} = \Lambda_\times$, given fixed timelike $\sqrt{-k^2}$, where the flow terminates at a maximal possible $\rho = \rhox$,

\textit{ii)} 	$\gamma$, as well as $\lambda_2$, must be real,  

\textit{iii)} 	$\gamma$ must be non-singular and must not break the unitarity bound $-\infty < \gamma \leq \nu + 1$,  

\textit{iv)}	$\gamma$ is expected to be monotonically increasing (never decreasing), $\frac{\partial\gamma}{\partial\rho} \geq 0$, from initial $\gamma(0) = 0$, hence $\gamma \geq 0$ for all $\rho$.

\subsection{Pure $AdS_{d+1}$}\label{Sec:AnomDimAdS}

As for our first example, let us find the anomalous dimension of a scalar operator dual to a massive scalar in pure $AdS_{d+1}$, with the metric in Poincar\'{e} coordinates given by equation \eqref{AdSMet1}. This example describes a large-$N$ theory conformal in its single-trace sector. For completeness, we repeat the metric here:
\begin{equation}\label{AdSMet1Rep}
ds^2 = \frac{1}{r^2} \left( -dt^2 + d\vec{x}^2_{d-1} +dr^2  \right).
\end{equation}
Writing $\lambda_2 (\rho) = \Delta_- + \gamma(\rho)$, the equation \eqref{DTRGEq} for the flow of $\gamma$ becomes
\begin{equation}\label{EqGamma1}
\rho \frac{\partial \gamma}{\partial \rho} = - \gamma \left( \gamma - 2 \nu \right) +  \rho^2 k^2.
\end{equation}
To see how condition \textit{iv)} applies, we analyse the case of a timelike physical momentum $k^2 < 0$. This means that $\rho^2 k^2 \leq 0$. Since $\rho \frac{\partial \gamma}{\partial \rho} \geq 0$, the only way for the right-hand side of \eqref{EqGamma1} to be non-negative is if $\gamma - 2 \nu$ remains sufficiently negative while $\rho$ increases. To see this, note that at $\rho=0$, $\gamma (\gamma - 2\nu) = 0$, as well as $\rho^2 k^2 = 0$. Now as $\rho$ increases the first term, $-\gamma(\gamma-2\nu) \geq 0$ grows larger until $\gamma > 2 \nu$. At that point, the overall sign of the first term flips and becomes negative. The second term, however, decreases monotonically into negative values and may quickly begin to dominate over the first term, running the entire right-hand-side of \eqref{EqGamma1} into negative values before $\gamma = 2\nu$. Function $\gamma(\rho)$ therefore reaches its maximal value $\gamma_\times$, at some $\rhox$, when $- \gamma_\times \left( \gamma_\times - 2 \nu \right) +  \rhox^2 k^2 = 0$. The two possible solutions, $\gamma_\times = \nu \pm \sqrt{ \nu^2 + \rhox^2 k^2 }$, can only be real and consistent with condition \textit{ii)} if $\rhox \sqrt{-k^2}  \leq \nu$. But given that we seek maximal $\rhox$, this inequality implies that $\rhox = \nu / \sqrt{-k^2}$. The largest possible value of $\gamma$, given some timelike momentum and $\nu$, is therefore $\gamma_\times = \nu$ irrespective of our choice of solution, confirming the renormalisation conditions specified by \textit{i)}. Despite this, the correct solution would be the one with the minus sign because at $\rhox = 0$, as required when $-k^2 \to \infty$, $\gamma_\times$ should vanish for it to be consistent with the initial condition \eqref{L2Def}. Only in this case can $\gamma$ and its derivative be continuous and non-singular. Furthermore, since the right-hand-side of \eqref{EqGamma1} vanishes at $\rhox>0$, we clearly have a maximum $\frac{\partial \gamma_\times}{\partial \rho} =0$ for any timelike operator momentum at the point where the RG flow terminates. Note that we did not have to impose the vanishing of the derivative, as dictated by the renormalisation condition \textit{i)} at $\rhox$ by hand, even though it followed from a general field theory analysis. It is automatically satisfied, as is \textit{iii)}, and we chose to start with condition \textit{iv)} to show the internal consistency of the four conditions.

Alternatively, we could have started with the renormalisation conditions \textit{i)}. Looking for the maximum would imply that $\gamma_\times = \nu$ at a maximal possible $\rhox$. It can then be seen that condition \textit{iv)} is also satisfied in order for $\gamma$ to obey \textit{ii)} and \textit{iii)}. This means that when \textit{ii)} and \textit{iii)} are enforced, \textit{i)} implies \textit{iv)}, but also \textit{iv)} implies \textit{i)}. Therefore, the two conditions are equivalent: $\textit{i)} \Leftrightarrow \textit{iv)}$.

\begin{figure}[!ht]
\centering
\includegraphics[scale=0.7]{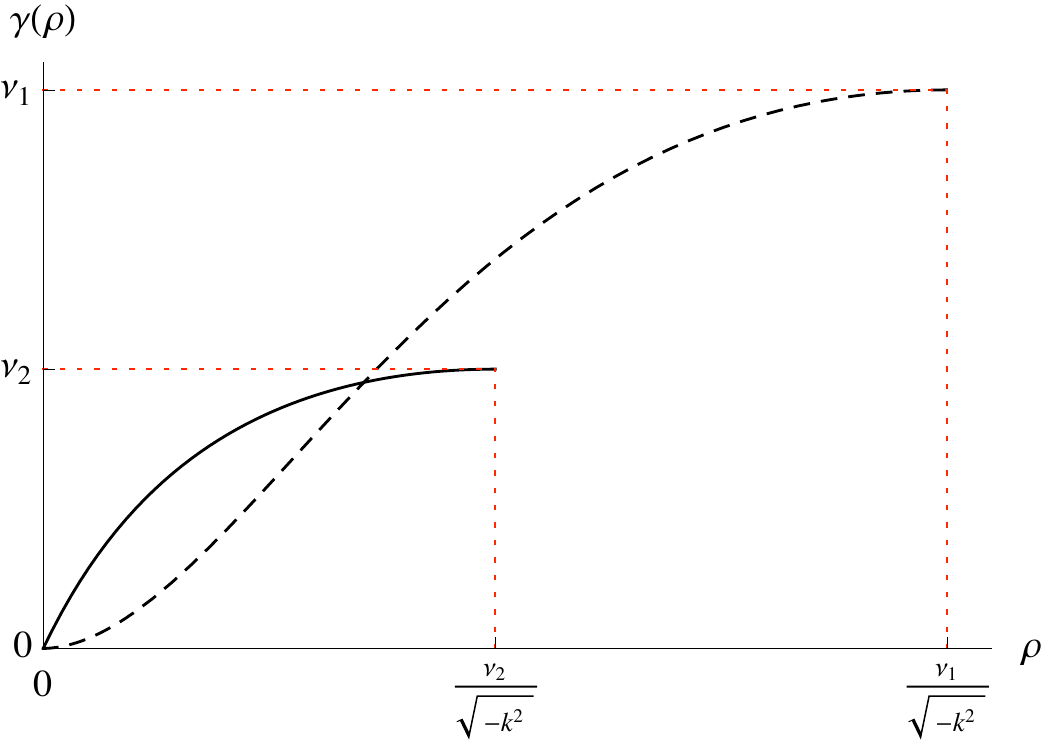}
\caption{Solutions \eqref{EtaAdSSol} for anomalous dimensions $\gamma$ with $\nu_1 = 1$ (dashed curve) and $\nu_2 = 0.5$ (solid curve).}
\label{fig:AdSGamma}
\end{figure}

The solution of equation \eqref{EqGamma1} can be found explicitly for the anomalous dimension with $\gamma=0$ at $k=0$. Written in terms of Bessel functions, it takes the form
\begin{multline}\label{EtaAdSSol}
\gamma(\rho)   =    \rho  \sqrt{-k^2} \left\{ \frac{  Y_{\nu -1}(\nu ) J_{\nu -1}( \rho  \sqrt{-k^2}) - J_{\nu -1}(\nu ) Y_{\nu -1}( \rho  \sqrt{-k^2})   }{  \left[ Y_{\nu -1}(\nu ) - Y_{\nu }(\nu )  \right]  J_{\nu }( \rho  \sqrt{-k^2}) -   \left[ J_{\nu -1}(\nu )  - J_{\nu }(\nu )  \right]  Y_{\nu }( \rho  \sqrt{-k^2})   } - \right.  \\
\left. - \frac{   Y_{\nu }(\nu ) J_{\nu -1}( \rho  \sqrt{-k^2})   -  J_{\nu }(\nu ) Y_{\nu -1}( \rho  \sqrt{-k^2})     }{  \left[ Y_{\nu -1}(\nu ) - Y_{\nu }(\nu )  \right]  J_{\nu }( \rho  \sqrt{-k^2}) -   \left[ J_{\nu -1}(\nu )  - J_{\nu }(\nu )  \right]  Y_{\nu }( \rho  \sqrt{-k^2})   }    \right\},
\end{multline}
with the range of $0 \leq \rho \leq \frac{\nu}{\sqrt{-k^2}}$. Plots of two different solutions with $\nu = 0.5$ and $\nu = 1$ are shown in Figure \ref{fig:AdSGamma}.

Note that the solution \eqref{EtaAdSSol} is well-defined for all real $\nu > 0$, above the Breitenlohner-Freedman bound \cite{Breitenlohner:1982jf}. For non-integer values of $\nu$, \eqref{EtaAdSSol} can be simplified to give
\begin{equation}\label{EtaAdSSolNonInt}
\gamma(\rho) =  \rho  \sqrt{-k^2} \frac{ \left[ J_{-\nu -1} (\nu ) + J_{-\nu } (\nu ) \right]  J_{\nu -1}( \rho  \sqrt{-k^2}) + \left[ J_{\nu -1}(\nu )-J_{\nu }(\nu ) \right] J_{1-\nu }( \rho  \sqrt{-k^2})}{\left[ J_{-\nu -1}(\nu )+J_{-\nu }(\nu ) \right] J_{\nu }( \rho  \sqrt{-k^2})+ \left[ J_{\nu }(\nu  ) -  J_{\nu -1}(\nu ) \right] J_{-\nu }( \rho  \sqrt{-k^2})}.
\end{equation}

The interpretation of how $\rho$ translates into the Wilsonian cut-off $\Lambda$ is now clear, as is the fact that the flow of $\gamma$ corresponds to the Wilsonian renormalisation group. If we reverse the argument, keep $\sqrt{-k^2}$ arbitrary, and insist on integrating out geometry between $0\leq\rho\leq\rhox$, then there is a limited interval of timelike momenta that operators can take after integration. The relation is 
\begin{equation}\label{KRhoxIneqAdS}
\sqrt{-k^2} \leq \nu / \rhox, 
\end{equation}
which implies the presence of a hard momentum cut-off on the brane side of holographic duality, induced by the sliding brane. It is precisely the hard Wilsonian UV cut-off with the Lorentzian signature $\sqrt{-k^2} \leq \Lambda$, that defines the energy scale up to which the effective field theory is valid. The situation is somewhat different from the Euclidean field theory case, and we are effectively integrating out energy-momentum modes in regions above floating hyperbolae in a light-cone diagram, down to asymptotically lightlike momenta. The lightlike $k^2 = 0$ can nevertheless not be reached and we need to treat that case separately. In real-time field theory, Lorentzian cut-offs always have a smooth form to ensure that the gauge invariance is preserved. Furthermore, a hard cut-off may not be sufficient to regularise the UV divergences due to the infinite volume of the energy-momentum space under the hyperbola. Nevertheless, it is natural that a Lorentzian Wilsonian cut-off should have a well-defined physical meaning, consistent with relativity. It suggests a re-ordering of physical phenomena according to their invariant length scales and implies a mixing of the usual Euclidean IR/UV separation of the energy scales. In contrast with the Lorentzian relativistic view on the renormalisation group scales, the Euclidean ordering of physical phenomena according to their length scales is motivated by locality in space.\footnote{We thank Janos Polonyi for ideas and discussions on various issues related to the interpretation of a Lorentzian cut-off.} A Lorentzian cut-off may be especially suitable for the investigation of non-perturbative theories and the operator-product renormalisation of local composite operators $\CO$, which are singlets under the gauge group. 

The same inequality, \eqref{KRhoxIneqAdS}, holds for any chosen momentum scale $k^2$ of an operator as well as any chosen scale $\rhox$, where we decide to terminate the integration. A scaling symmetry between $k$ and $\rho$ is apparent from the evolution equation \eqref{EqGamma1}, which is invariant under simultaneous 
\begin{equation}\label{ScalingEqGamma1}
\rho \to a \rho ~\text{and}~ \sqrt{-k^2} \to \sqrt{-k^2} / a,~ \text{with constant}~ a.
\end{equation}
The analysis therefore provides an exact functional dependence for the correspondence between parameters describing the bulk physics and their dual Wilsonian UV cut-off $\Lambda(\rho,d,m,...)$ for all $\rho$ and $k$. It is also important for this identification that the operator $\rho \partial / \partial\rho$ is invariant under $\rho \to a \rho$ for constant $a$. Hence the boundary energy scale is 
\begin{equation}
\Lambda = \nu / \rho.
\end{equation}
$\Lambda_\times = \nu / \rhox$ is then the lowest possible scale down to which we can integrate from $\Lambda \to \infty$, given some momentum $k$ at which we wish to evaluate the operator. Energy scale $\Lambda$ must be real and positive, so $\nu$ must also be real and positive. This is again ensured by the Breitenlohner-Freedman bound \cite{Breitenlohner:1982jf}, which we analogously required for a well-defined range of $\rho$. It is important to note that the constant of proportionality $\nu$ is merely a result of the bulk coordinates we used in \eqref{AdSMet1Rep} to establish the dependence between that particular bulk space and its boundary dual. It also clearly signals the expected probe dependence on the bulk scalar mass. We could easily redefine $r \to \nu r$ to give us a metric 
\begin{equation}\label{AdSMet2}
ds^2 = \frac{1}{\nu^2 r^2} \left( -dt^2 + d\vec{x}^2_{d-1} \right) + \frac{dr^2}{r^2}.
\end{equation} 
In this background, we obtain $\sqrt{-k^2} \rhox \leq 1$ and hence $\Lambda=1/\rho$. For a general Poincar\'{e}-like AdS chart, we can therefore conclude that the Wilsonian energy scale of the cut-off in a boundary theory is related to the radial bulk coordinate by
\begin{equation}\label{LambdaBulkAdSRelat}
\Lambda = \frac{C(d,m,...)}{r},
\end{equation} 
where $C$ is a constant, which depends on the bulk quantities describing the background metric and can be found exactly following the above procedure. Our analysis is thus consistent with the long anticipated relationship $\Lambda \propto 1 / r$ \cite{Susskind:1998dq,Peet:1998wn}. In addition, it also uniquely determines the proportionality constant for a given pair of holographically dual quantities, i.e. the composite scalar operator $\CO$ and its dual massive scalar field probe in the AdS background. 

At the Breitenlohner-Freedman bound with $\nu=0$, and for arbitrary momentum, the RG flow analysis breaks down unless $\rho_\times = 0$. This is also apparent from the metric \eqref{AdSMet2} which is singular at $\nu=0$. The only way to have an RG flow compatible with such an operator at the UV fixed point is when $\CO$ has lightlike on-shell momentum $k^2=0$. The solution is still $\gamma(\rho)=0$ and the anomalous dimension does not run, but it is well-defined for all $\rho$.

We need to consider the lightlike (or vacuum) $k^2 = 0$ case separately. Equation \eqref{EqGamma1} drastically simplifies for all operator dimensions. To satisfy the monotonicity condition \textit{iv)}, the anomalous dimension must behave as $\gamma_\times \to 2 \nu$ when $\rhox \to \infty$. The solution of equation \eqref{EqGamma1} that satisfies the required conditions is then
\begin{equation}\label{EtaAdSSolLight}
\gamma(\rho) = \frac{2 \nu \chi \rho^{2\nu}}{1 + \chi \rho^{2\nu}},
\end{equation}
as previously found by \cite{Faulkner:2010jy,Vecchi:2010dd}. In a timelike scenario considered above, it is not possible to flow to $\gamma = 2 \nu$ without violating condition \textit{iii)}, as there is always a singularity between $\gamma(0) = 0$ and $\gamma=2\nu$, unless $k=0$. This is precisely what prevents us from having a continuous, well-defined limit between timelike and lightlike cases. Constant $\chi$ cannot be determined from the boundary conditions we set, but needs to be matched with the normalisation of the corresponding two-point correlator.

For spacelike momenta $k^2 > 0$, the energy scale becomes pure imaginary. Having found the radial coordinate to be proportional to the energy of the dual theory, it is therefore natural to take $r \to i r$, consistent with rescaling symmetry relations \eqref{ScalingEqGamma1}. The analysis of \eqref{EqGamma1} then goes through in exactly the same way as for timelike momenta. We obtain $k^2 \leq \Lambda_\times = \nu / \rho_\times$ and $\Lambda = C / r$.

\subsection{GPPZ flow from $\CN=4$ to $\CN=1$ with a mass deformation and a mass gap}\label{Sec:AnomDimGPPZ}

The GPPZ flow \cite{Girardello:1999bd} describes a flow from the $\CN=4$ theory in the UV to the  $\CN=1$ in the IR. This is achieved by deforming the $\CN=4$ theory with a relevant mass deformation. An $\CN=4$ vector multiplet in the adjoint is equivalent to three $\CN=1$ chiral multiplets in the adjoint. Giving mass to one of the $\CN=1$ mutiplets provides the desired deformation. The theory can flow in different ways depending on which components of the deformation are kept non-zero. They are described in \cite{Girardello:1998pd,Girardello:1999bd,Porrati:2000nb,Porrati:2000ha}. We will focus on the case that includes the $\CN=1$ confinement, suppressing the supergravity singlet dual to a bilinear gaugino term responsible for gaugino condensate. The bulk supergravity for this construction consists of type IIB scalar modes deforming the original $AdS_5 \times S^5$ metric. This $10$-dimensional type IIB theory is then truncated on $S^5$ to give a $5$-dimensional $\CN=8$ supergravity with 42 scalars. The scalars transform as \textbf{1}, \textbf{20} and \textbf{10} under the $\CN=4$, $SU(4)_R$, R-symmetry. The masses of these fields are $m^2=0$, $-4$ and $-3$, respectively. The GPPZ flow then describes the metric deformation resulting from the backreaction with scalars of $m^2=-3$. This corresponds to a dual deformation of dimension $3$ by the scalar operator in $4$ spacetime dimensions, which can be identified as the fermion bi-linear operator with the coupling constant of mass dimension $1$. The \textbf{10} of $SU(4)$ decomposes into $\textbf{1}+\textbf{6}+\textbf{3}$ under $SU(3) \times U(1)$, of which we only keep the \textbf{6}, which is responsible for the fermion bilinear mass term. The singlet would give rise to a gaugino condensate \cite{Girardello:1999bd}. 

We further truncate the theory to only account for the large-$N$-dominant effective double-trace deformations and consider the background as static. The supergravity scalar potential is then $V(\Phi) = -3 - \frac{3}{2} \Phi^2 + \CO(\Phi^4)$, where the constant first term plays no role in classical equations of motion. This results in the same scalar theory we have been considering so far with mass $m^2 = -3$ and a $5$-dimensional metric
\begin{equation}\label{GPPZMetric}
ds^2 = \frac{1}{r^2} \left( 1-\frac{r^2}{r_1^2}  \right)  \left( -dt^2 + d\vec{x}^2  \right) + \frac{dr^2}{r^2}.  
\end{equation}
By $r_1$ we denote the radial position where the flow terminates at a metric singularity. Note that this metric is simply obtained from its original form in \cite{Girardello:1999bd} by defining $y=-\ln r$. The metric interpolates between an $AdS_5$ space near $r=0$ with a conformal group $SO(4,2)$ and a $4$-dimensional Poincar\'{e} group near $r=r_1$. Taking $r_1 \to \infty$ would give an interpolation into the extreme IR. 

We can now analyse the renormalisation group flow of a scalar operator deformed by both effective double-trace and a relevant mass deformation. The mass deformation is provided by the GPPZ flow of the supergravity metric through its $r$ dependence and the double-trace deformation is ensured by the quadratic term in the bulk scalar potential. Writing as before $\lambda_2 = 2-\nu+\gamma$, the renormalisation group equation \eqref{DTRGEq} for $\gamma$ in background \eqref{GPPZMetric} becomes
\begin{equation}\label{EqGamma3}
\rho  \frac{\partial \gamma}{\partial \rho} \left(1 - \frac{\rho^2}{r_1^2} \right)^2 = - \gamma \left( \gamma - 2 \nu \right) \left(1 - \frac{\rho^2}{r_1^2} \right)^2 + \frac{4 \rho^2}{r_1^2}  \left( 2- \nu + \gamma \right)  \left(1 - \frac{\rho^2}{r_1^2} \right) +  \rho^2 k^2 \left(1 - \frac{\rho^2}{r_1^2} \right).
\end{equation}
We need to be careful not to divide by zero in case $\rho \to r_1$, so we preserved the original form of the RG equation \ref{HFlowEq}. In the case of the GPPZ flow, $\nu = 1$. We may, however, keep $\nu$, as well as $m$ general for now, restricting it only to relevant and marginal deformations with $\nu \leq 2$ so that $(2-\nu+\gamma) \geq 0$. An irrelevant deformation could anyhow not be described by a factor like $\left(1 - \rho^2 / r_1^2 \right)$, which breaks the UV symmetries with increasing intensity as the flow approaches the IR regime. We use the same reasoning as in the pure AdS case in section \ref{Sec:AnomDimAdS} to determine whether the flow can terminate at a fixed point given some timelike momentum $k^2 < 0$. Given that the pure AdS scenario corresponds to the undeformed conformal $\CN=4$ theory, it is especially interesting to compare its RG flow with the RG flow in this section. 

Equation \eqref{EqGamma3} can be solved analytically. Its complicated explicit solution is, however, not too illuminating and will not be presented. Qualitatively, its graph behaves as that of the solution \eqref{EtaAdSSol} in pure AdS, which we plotted in Figure \ref{fig:AdSGamma}. In the present case, we will only look for the functional dependence of $\Lambda$ on the bulk. Starting as in section \ref{Sec:AnomDimAdS} with \textit{iv)} for a monotonically increasing, positive and non-singular $\gamma$, the only term which can run the right hand side of \eqref{EqGamma3} into negative values for $\gamma < 2 \nu$ is $\rho^2 k^2$. As before, to find the maximal $\gamma_\times$ at $\rhox$, the right-hand side of \eqref{EqGamma3} has to vanish, implying the renormalisation condition \textit{i)}. The solutions are
\begin{equation}
\gamma_\times = \nu + \frac{2 \rhox^2}{r_1^2 - \rhox^2} \pm \frac{r_1^2}{r_1^2 - \rhox^2} \sqrt{  \nu^2  \left(1 - \frac{\rhox^2}{r_1^2} \right)^2  + \rhox^2 k^2  \left(1 - \frac{\rhox^2}{r_1^2} \right) + 8 \frac{\rhox^2}{r_1^2} \left(1 - \frac{\rhox^2}{r_1^2} \right) + 4 \frac{\rhox^4}{r_1^4} }.
\end{equation} 
For $\gamma_\times$ to be real when $\rhox$ is maximal (condition \textit{ii)}), the expression under the square-root needs to be non-negative. The term $\sim \rhox^2 k^2 < 0$ will however cause the expression to inevitably flow towards $0$. We could, at this point, impose a condition whereby the entire expression under the square-root would be non-negative and thus vanishing at maximal $\rhox$, given some $k$. This is, however, insufficient. For $\rhox > r_1/ \sqrt{3}$, $\gamma_\times$ would break the unitarity condition \textit{iii)}. Another problem is that in the case of $\rhox = r_1$, despite the vanishing square-root, the second term in $\gamma_\times$, $2 \rhox^2 / \left( r_1^2 - \rhox^2 \right)$, would blow up. This would thus also violate the non-singularity condition, so again \textit{iii)}. To remedy these problems, we, as before, select the solution with the minus sign and impose
\begin{equation}\label{KRhoxIneqGPPZ1}
\nu^2  \left(1 - \frac{\rhox^2}{r_1^2} \right)  + \rhox^2 k^2 + 8 \frac{\rhox^2}{r_1^2}  \geq 0,
\end{equation}
which at maximal $\rhox$ turns into an equality and enables the $4 \rhox^4 / r_1^4$ term under the square-root to cancel the diverging second term in $\gamma_\times$ as $\rhox \to r_1$. We therefore arrive at condition \textit{i)}, $\gamma_\times = \nu$ and $\partial_\rho \gamma_\times = 0$, at maximal $\rhox$. This is consistent with \cite{Porrati:2000nb}, where it was shown that the GPPZ construction indeed flows between two fixed points. 

Alternatively, as before, we could have simply imposed condition \textit{i)} and looked for maximal possible $\rho$ without violating \textit{ii)} and \textit{iii)}, arriving at the same conclusion. This would then imply \textit{iv)}.

There is a scaling symmetry in equation \eqref{EqGamma3}, equivalent to \eqref{ScalingEqGamma1}, but with an extra rescaling of $r_1$:
\begin{equation}\label{ScalingEqGamma2}
\rho \to a \rho, ~ r_1 \to a r_1  ~ \text{and}~ \sqrt{-k^2} \to \sqrt{-k^2} / a,~ \text{with constant}~ a.
\end{equation}

We can use the same analysis at any chosen momentum. Rewriting inequality \eqref{KRhoxIneqGPPZ1} as a momentum bound then determines the exact functional dependence of the Wilsonian cut-off $\Lambda$ on the bulk quantities. Equivalently, we could determine the range of allowed consecutive momentum shell integrations given some operator momentum we wish to probe. The relation is
\begin{equation}\label{KRhoxIneqGPPZ2}
-k^2 \leq \frac{\nu^2}{\rhox^2} + \frac{4-m^2}{r_1^2} = \Lambda^2_{\CN = 4 \to 1},
\end{equation}
where $4-m^2=8-\nu^2$, and in the case of the GPPZ flow $- k^2 \leq 1/\rhox^2 + 7/r_1^2$. The interpretation of \eqref{KRhoxIneqGPPZ2} clearly shows the interplay of two scales: the Wilsonian energy scale $\Lambda^2_{\CN=4} = \nu^2 / \rho^2$ of the undeformed $\CN = 4$ theory and the fixed ($\rho$-independent) mass deformation scale 
\begin{equation}\label{MassDefScale}
\Lambda_{\CM}^2 = \frac{4-m^2}{r_1^2}, 
\end{equation}
which is independent of $\Lambda_{\CN=4}$. $\Lambda_\CM$ is a constant scale and can be tuned to a desired fixed value by setting $m$ and $r_1$. In a realistic scenario, however, $r_1$ should be very large so that the flow terminates in the IR. In the UV regime with $\rho << r_1$, $\Lambda_{\CN=4}$ dominates the cut-off of the deformed theory. Then flowing into the IR, as expected from a relevant deformation, $\Lambda_{\CM}$ becomes increasingly important. The square of the Wilsonian scale of the deformed theory can simply be written as the sum of squares of both scales present in the theory
\begin{equation}\label{GPPZasTwoScales}
\Lambda^2_{\CN=4\to1} = \Lambda^2_{\CN=4} + \Lambda^2_\CM.
\end{equation} 

It is interesting to further study the limit $\rho \to r_1$. \eqref{KRhoxIneqGPPZ2} implies $\sqrt{-k^2} \leq 2 \sqrt{2} / r_1$ independently of the value $\nu$. However, if we investigate \eqref{EqGamma3} directly, terms with $\partial\gamma / \partial \rho$ and $\gamma(\gamma-2\nu)$ vanish faster than the remaining two terms on the right-hand-side of the equation. We of course have to assume well-behaved, non singular $\gamma$ and $\partial_\rho \gamma$ consistent with \textit{iii)}. Equation \eqref{EqGamma3} is then solved at the linear order in $(r_1 - \rho)$ with $\gamma_\times = \nu$, by turning \eqref{KRhoxIneqGPPZ2} into an equality:
\begin{equation}\label{SolGPPZatR1}
\sqrt{-k^2} = 2 \sqrt{2} / r_1.
\end{equation}
This indicates that momenta lower than the cut-off scale $\Lambda_{\CN=4\to1}$ are unstable at the endpoint of the flow. The only other solutions existing at $\rhox = r_1$ that we can find to \eqref{EqGamma3} with $\partial_\rho \gamma_\times = 0$ must have either $\gamma = 0$ or $2\nu$, similar to the $k^2=0$ case in pure AdS. Momentum then takes two possible values
\begin{align}\label{SolGPPZatR12}
\sqrt{-k^2} = \frac{2 \sqrt{2-\nu}}{r_1} ,~\text{for} ~ \gamma = 0& & \text{and}  & &\sqrt{-k^2} = \frac{2 \sqrt{2+\nu}}{r_1}, ~ \text{for} ~ \gamma = 2 \nu.  
\end{align}
To satisfy the unitarity bound \textit{iii)}, $\gamma \leq \nu + 1$, the second solution is permitted only when $\nu \leq 1$, which is saturated in the GPPZ flow, thus allowing for all three momenta: $2\sqrt{2} / r_1$ and $2\sqrt{2 \pm 1} / r_1$ at $\rhox=r_1$. The case of $k^2 = 0$ with $\partial_\rho \gamma_\times = 0$ would give $\gamma_\times = \nu - 2$ solution at $\rho \to r_1$, as we have to permit such operators to flow until the end. In the GPPZ construction, it would equal $\gamma = -1$ at $\rho = r_1$. However, as we discussed above, it does not make sense to impose $\partial_\rho \gamma = 0$ for $k^2 = 0$ before $\rho \to \infty$. In any case, we subtract off the vacuum from $\langle\CO(k)\CO(-k)\rangle$, so it is only important that the GPPZ vacuum is stable. 

The discreteness of possible momenta above the vacuum is a clear indicator that there is a mass gap present in the theory. There are three discrete states in the spectrum. This is consistent with the confining nature of the GPPZ flow already discussed in \cite{Girardello:1999bd}.    

The analysis in case of spacelike momenta goes through as above, after we rotate the two energy scales $r \to i r$ and $r_1 \to i r_1$, as indicated by rescaling symmetries \eqref{ScalingEqGamma2}. 

\subsection{Black brane in $AdS_{d+1}$ with thermal and density deformations, a mass gap and the emergent infra-red CFT scaling}\label{Sec:AnomDimAdST}

We now turn our attention to backgrounds giving non-zero temperature $T$ to their field theory duals. The example we consider first is the metric of a black brane in a $(d+1)$-dimensional AdS space of form \eqref{BBAdSMetric0}, given by
\begin{equation}\label{BBAdSMetric}
ds^2 = \frac{1}{r^2} \left( - f(r) dt^2 + d\vec{x}^2_{d-1} + \frac{dr^2}{f(r)} \right),~\text{with}~f(r)=1 - \frac{r^d}{r_0^d}.
\end{equation}
Using \eqref{L2DefT}, $\lambda_2(\rho) = \sqrt{f(\rho)} \left( \frac{d}{2} - \nu + \gamma(\rho) \right)$, the renormalisation group flow equation \eqref{DTRGEq} for $\gamma$ becomes
\begin{equation}\label{EqGamma2}
f \rho \frac{\partial \gamma}{\partial\rho} = - \gamma \left( \gamma - 2 \nu \right) + \frac{\rho^d}{r_0^d} \left( \frac{d}{2} - \nu + \gamma\right)^2 - \frac{ \rho^2 \omega^2}{f} + \rho^2 \vec{k}^2.
\end{equation}
In analogy with \eqref{ScalingEqGamma1} and \eqref{ScalingEqGamma2}, there is a scaling symmetry,
\begin{equation}\label{ScalingEqGamma3}
\rho \to a \rho, ~ r_0 \to a r_0  , ~ \omega \to \omega / a ~ \text{and}~ |\vec{k}| \to |\vec{k}| / a ~ \text{with constant}~ a.
\end{equation}
We can use the same procedure exactly as in sections \ref{Sec:AnomDimAdS} and \ref{Sec:AnomDimGPPZ} to extract the dependence of the Wilsonian cut-off scale on the bulk, consistent with conditions \textit{i)} - \textit{iv)}. First, 
\begin{equation}
\gamma_\times = \nu + \frac{d}{2} \frac{\rhox^d}{r_0^d - \rhox^d} - \frac{r_0^d}{r_0^d - \rhox^d} \sqrt{ \frac{d^2}{4} \frac{\rhox^d}{r_0^d}      +    \nu^2 \left(1 -  \frac{\rhox^d}{r_0^d} \right)  - \rhox^2 \omega^2 + \rhox^2 \vec{k}^2 \left(1 -  \frac{\rhox^d}{r_0^d} \right)  }.
\end{equation}
The expression is qualitatively similar to the case of the GPPZ flow in section \ref{Sec:AnomDimGPPZ}. Following the same considerations which satisfy conditions \textit{ii)} and \textit{iii)} leads to $\gamma_\times = \nu$ and an inequality
\begin{equation}
\frac{d^2}{4} \frac{\rhox^d}{r_0^d}     +    \nu^2 f (\rhox)  - \rhox^2 \omega^2  + \rhox^2 \vec{k}^2 f(\rhox)  \geq  \frac{d^2}{4} \frac{ \rhox^{2d}}{ r_0^{2d}},
\end{equation}
which can be recast as an energy-momentum bound 
\begin{equation}\label{KRhoxIneqAdST}
\frac{\omega^2}{f(\rhox)} - \vec{k}^2 \leq \frac{\nu^2}{\rhox^2} + \frac{d^2}{4} \frac{\rhox^{d-2}}{r_0^d} .
\end{equation}
It is clear from the left-hand side of \eqref{KRhoxIneqAdST} that, as a result of temperature, energy and momentum scale differently. The scaling difference comes from the thermal factor $f(r)$, which completely breaks the Lorentz invariance of the boundary theory away from the UV fixed point with the $\CN=4$ theory. This is expected to happen in any thermal field theory. The Hawking temperature is 
\begin{equation}\label{T}
T = \frac{d}{4\pi r_0},
\end{equation}
which enables us to rewrite \eqref{KRhoxIneqAdST} as
\begin{equation}\label{KRhoxIneqAdST2}
\frac{\omega^2}{f(\Lambda, T)} - \vec{k}^2 \leq  \Lambda^2 +  \frac{d^2}{4 \nu^2}  \left( \frac{4\pi\nu}{d} \right)^d \left( \frac{T}{\Lambda} \right)^d \Lambda^2 \equiv \Lambda^2_{\CN=4,T},  
\end{equation}
with
\begin{equation}
f (\Lambda,T) = 1 - \left( \frac{4\pi \nu}{d} \right)^d  \left( \frac{T}{\Lambda} \right)^d , 
\end{equation}
having used $\Lambda = \nu / \rho$ with the fact that we can apply the same analysis to any value of energy-momentum, thus removing the need to only consider the minimal cut-off $\Lambda_\times$. The Wilsonian scale has the same bulk dependence throughout the flow. We can then, for the same reason, define the thermal scale depending on $\Lambda$, and not only $\Lambda_\times$, as
\begin{equation}\label{Tscale}
\Lambda^2_T \left( \Lambda, \frac{T}{\Lambda } \right) =  \frac{d^2}{4 \nu^2}  \left( \frac{4\pi\nu}{d} \right)^d \left( \frac{T}{\Lambda} \right)^d \Lambda^2, 
\end{equation}
where we have been using the abbreviation $\Lambda = \Lambda_{\CN=4}$, for the Wilsonian scale at $T=0$. In $d=4$, the thermal scale is $\Lambda_T^2 = 4 \pi^2 \left(\pi \nu \Lambda \right)^2 (T / \Lambda)^4$. We see that, as with the GPPZ case in equation \eqref{GPPZasTwoScales}, the two relevant scales, the undeformed ($T=0$) Wilsonian scale and the temperature deformation add quadratically to give the new Wilsonian cut-off of the $\CN=4$ theory with the temperature deformation breaking Lorentz invariance:
\begin{equation}\label{AdSTasTwoScales}
\Lambda^2_{\CN=4,T} = \Lambda^2_{\CN=4} + \Lambda^2_T.
\end{equation}

All temperature-dependent terms that appear in the final cut-off scale relation \eqref{KRhoxIneqAdST2} appear with $T$ to the power of the boundary field theory dimension, $T^d$. That is in the thermal factor $f( T/\Lambda )$, as well as in the square of the thermal scale $\Lambda^2_T (\Lambda, T/\Lambda)$. $\Lambda^2_T$ depends on both the original undeformed scale $\Lambda$ and the dimensionless $T/\Lambda$. We will see the same dimensional power-law behaviour of temperature, $T^d$, in all thermal cases we consider.

From relation \eqref{KRhoxIneqAdST2} we see that the renormalisation group procedure particularly restricts the energy $\omega$. The spatial momentum $|\vec{k}|$ can be arbitrarily large without violating the bound because of a minus sign in front of it. In fact, the larger the momentum, the larger range of values $\omega$ can take. In the limit of $\rhox \to r_0$, energy becomes completely suppressed and must go to zero at least as fast as $f(\rho)$. The restrictions posed on $\omega$ are not surprising as temperature is an energy scale. We also see that the effect of $\Lambda_T$ becomes increasingly important when we run into the IR because $\Lambda_T \propto \Lambda^{2-d}$, giving it an overall negative power in a dimension greater than two. This is consistent with temperature being an IR scale. Taking the limit of $\rhox \to r_0$ in \eqref{KRhoxIneqAdST}, we get $-\vec{k}^2 \leq \frac{1}{r_0^2} \left(\nu^2 + \frac{d^2}{4} \right)$. However, analogously to the GPPZ case in section \ref{Sec:AnomDimGPPZ}, the renormalisation group equation \eqref{EqGamma2} at the horizon with $\omega = 0$ strictly enforces the equality 
\begin{equation}\label{MassGap4T}
-\vec{k}^2 = \frac{1}{r_0^2} \left(\nu^2 + \frac{d^2}{4} \right), 
\end{equation}
for a well-behaved $\partial_\rho \gamma_\times$ and $\gamma_\times = \nu$. The mass of the momentum mode \eqref{MassGap4T} is then $M^2 = - \vec{k}^2$, which clearly shows the emergence of a mass gap with only one permitted value of $M$ above the vacuum. Similarly, we can set momentum to zero, i.e. $\vec k =0$, scale energy as $\omega^2 = \tilde\omega^2 f(\rho)$, use $\gamma_\times = \nu$ and expand \eqref{EqGamma2} to first-order in the near-horizon limit. We then again recover the same gap in the energy spectrum of $\tilde\omega$ as we saw in the spectrum of momentum modes in \eqref{MassGap4T}. The existence of a mass gap in the $\CN=4$ theory at non-zero temperature was shown from an AdS/CFT calculation in \cite{Witten:1998zw}.

In imaginary-time formulation, we take $\omega \to i \omega$, $r\to i r$ and $r_0 \to i r_0$. After this transformation, the analysis follows through exactly as above. The only difference is that there is now a plus sign in front of $\vec{k}^2$ in the relation \eqref{KRhoxIneqAdST2}. After we have lost the Lorentzian nature of the original undeformed theory, momentum becomes just as suppressed as energy.

Lastly, we turn our attention to dimensional reduction of the emergent IR CFT scaling when the renormalisation group flow reaches the horizon $\rhox \to r_0$. In this limit, the thermal scale becomes
\begin{equation}\label{ThermalHorizonScale}
\lim_{\Lambda \to 4\pi\nu T / d} \Lambda^2_{T} = 4 \pi^2 T^2,
\end{equation}
where $\Lambda$ is the undeformed Wilsonian scale. Of course, this should be a fixed point of the deformed theory, as $\partial_\rho \gamma_\times = 0$ at the horizon. If we now look at the thermal scale for an arbitrary value of the undeformed $\Lambda(\rho)$ in $d=2$ dimensions, we notice that just, as in \eqref{ThermalHorizonScale},
\begin{equation}
\Lambda^2_T = 4 \pi^2 T^2, ~\text{when} ~ d=2,
\end{equation}
thus giving us a relation
\begin{equation}\label{ThermalScalesEquiv}
\Lambda^2_{T} (\Lambda = 4\pi\nu T / d , d)  = \Lambda^2_T  (\Lambda, d=2).
\end{equation}
It is essential to note that from the point of view of the bulk, the Hawking temperature $T$ is still a function of the number dimensions $d$. If we considered a fixed position of the horizon, then equality in \eqref{ThermalScalesEquiv} would be incorrect. But we know that physics in the bulk does not change, and nor does its holographic dual if we adjust the position of the horizon. The Ricci scalar curvature of the black brane \eqref{BBAdSMetric} does not depend on the position of the horizon $r_0$, which means that we can tune it to give us any temperature we want. Such tuning does not violate the validity of the classical gravity description of the bulk physics, which is essential for the application of holographic duality. The saddle point approximation of string theory remains applicable. From the point of view of the boundary QFT, temperature can be thought of as an adjustable parameter that can be set to any value independently of the number of dimensions. We are therefore really comparing two theories in a different number of dimensions, for which the two temperatures have been adjusted to be equal by adjusting the horizons. Thermal scales are then equal in the $(d=2)$ two-dimensional boundary theory with an arbitrary undeformed cut-off $\Lambda$, and in the horizon limit in an arbitrary number of dimensions. The two-dimensional boundary theory is conformal at all scales, as $\Lambda_T$ can be tuned to any fixed value, resulting from its $\Lambda$ independence. The emergent $d$-dimensional CFT therefore behaves near the horizon, in terms of scales, as a two-dimensional conformal field theory. This result is highly reminiscent of the work by Carlip \cite{hep-th/9812013} and Solodukhin \cite{hep-th/9812056}. They showed that with an appropriate choice of boundary conditions, extremal black hole horizons in any number of dimensions gave rise to an emergent two-dimensional Virasoro algebra of the asymptotic symmetries. Physics of an arbitrary black hole horizon could therefore be described by a two-dimensional CFT.\footnote{We thank Mukund Rangamani for bringing the work of Carlip and Solodukhin to our attention.}

Similar behaviour was discussed in the context of AdS/CFT in the case of an extremal charged black hole in $AdS_{d+1}$ \cite{Faulkner:2009wj,Faulkner:2010jy}. There, the metric takes exactly the same form as \eqref{BBAdSMetric}, with a different factor, $f(r) = 1 + Q^2  r^{2 d - 2} - M  r^d$. Dimensional reduction of the IR CFT in the charged case is apparent from the form of the metric. The function $f(r)$ has a double zero in the horizon limit, making the RG flow an interpolation between $AdS_{d+1}$ at $r \to 0$ and $AdS_2 \times \mathbb{R}^{d-1}$ at $r \to r_0$. The emergent CFT therefore appears to be $(0+1)$-dimensional, as opposed to ($1+1$) in our case. It was, however, argued in \cite{arXiv:0901.1677} and discussed in \cite{Faulkner:2009wj} that the near-horizon region of $AdS_2$ could be understood as being described by a single copy of a two-dimensional chiral Virasoro algebra of the asymptotic symmetries. The simple metric interpolation between two AdS spaces in the charged case does not occur in the thermal case, \eqref{BBAdSMetric}, since $f(r)$ only has a single zero at the horizon. Nevertheless, we find a similar emergent feature in the scaling of the theory.

To compare the charged case that gives rise to finite boundary density with the thermal case, we can use the generalised result for the Wilsonian momentum bound, \eqref{GenBound}, which we will derive in section \ref{Sec:AnomDimGen}. In this example, we see that we do not exactly recover the same behaviour of the dimensional reduction in scales, as in the purely thermal case. Using $f(r) = 1 + Q^2  r^{2 d - 2} - M  r^d$ and $h(r)=1$ in \eqref{GenBound}, we immediately find 
\begin{equation}\label{ChargedBnd}
\frac{\omega^2}{f} - \vec{k}^2 \leq \frac{\nu^2}{\rho^2} + \frac{d}{4} \rho^{d-2} \left(d M  - \left(3d-4\right) Q^2 \rho^{d-2} \right).
\end{equation}
The Hawking temperature for the charged metric is 
\begin{equation}\label{ChargedT}
T = \frac{d}{4\pi r_0} \left( 1 - \frac{(d-2)}{d} Q^2 r_0^{2d-2} \right),
\end{equation}
where $r_0$ is the position of the horizon, which is the smallest positive solution of $f(r_0) = 0$ and therefore also of
\begin{equation}\label{ChargedM}
M = \frac{1}{r_0^d} + Q^2 r_0^{d-2}.
\end{equation}
Using \eqref{ChargedT} and \eqref{ChargedM}, we can take the horizon limit, $\rho \to r_0$, of the Wilsonian bound \eqref{ChargedBnd}. The right-hand side of it becomes
\begin{equation}\label{ChargedBnd2}
\Lambda^2 + 4 \pi^2 T^2 - \frac{(d-2)^2}{4 r_0^2} Q^4 r_0^{4d-4} = \Lambda^2 + 4\pi^2 \left[ T^2 - \left(T - T_0 \right)^2 \right],
\end{equation}
with the usual $\Lambda = \nu / \rho$. We denoted the temperature at zero charge, $Q^2 = 0$, by $T_0 = \frac{d}{4\pi r_0}$. $T_0$ is therefore the temperature of the boundary theory at zero density. In two boundary dimensions, $d=2$, $T - T_0 = 0$ for all $\rho$. We therefore find the same conformal Wilsonian bound in the charged $d=2$ case, as in the purely massive (thermal) case, $\Lambda^2 + 4\pi^2 T^2$. This does not, however, equal \eqref{ChargedBnd2} for the horizon-limit bound in an arbitrary number of dimensions, unless $Q=0$ and we recover the purely thermal theory. We see a mixing between two temperature scales in the problem, one at zero and another at non-zero density of the boundary system.

\subsection{Near-extremal M2 and M5 branes}\label{Sec:AnomDimM}
\subsubsection{M2 brane}\label{Sec:M2sec}

We begin by writing down the metric of a non-extremal M2 brane by neglecting the spherical $d\Omega_7^2$ part,
\begin{equation}\label{M2Metric}
ds^2 = H(r)^{-2/3} \left( -f(r) dt^2 + d\vec{x}^2_2  \right) + H(r)^{1/3} \frac{dr^2}{f(r)},
\end{equation}
where $H(r) = 1 + R^6 / r^6$ and $f(r) = 1 - r_0^6 / r^6$. We take the near-extremal limit $r \ll R$, set $R=1$ and use $u = r_0^2 /r^2$ to find
\begin{equation}\label{NEM2Metric}
ds^2 = \frac{r^4_0}{u^2} \left( - f(u) dt^2 + d\vec{x}^2_2 \right) + \frac{du^2}{4 f(u) u^2},
\end{equation}
with $f(u) = 1 - u^3$. The horizon is at $u=1$. We further use a variable $q$, which is related to $u$ by $u = 2r_0^2 q$, to rewrite \eqref{NEM2Metric} in the form
\begin{equation}\label{NEM2Metric2}
ds^2 = \frac{1}{4 q^2} \left( - f(q) dt^2 + d\vec{x}^2_2 + \frac{dq^2}{f(q)} \right), 
\end{equation}
where $ f(q) = 1 - \frac{q^3}{q^3_0}$ and $q_0 = \frac{1}{2 r_0^2}$. This redefinition is necessary in order to avoid numerical factors in $G_{rr}$, which would change the boundary conditions for the anomalous dimension and introduce additional complications. The overall numerical rescaling factor of $1/4$ in \eqref{NEM2Metric2} is, however, irrelevant and can be absorbed into the AdS radius $R$, which we then reset to one. The renormalisation group equation \eqref{DTRGEq} for the anomalous dimension is itself invariant under an overall multiplication of the metric by a number. To see this, imagine rescaling a metric $ds^2 \to ds^2 / c^2$. Using \eqref{L2DefGrr} would mean that $\lambda_2 \propto c (\Delta_- + \gamma)$. The relation between the scalar mass and the dual operator dimension, which can be derived from the scalar equation of motion, would also be modified to $m^2 \propto - c^2 \Delta_- \Delta_+$. Therefore, each term in \eqref{DTRGEq} would become proportional to $c^2$ and we could cancel it out. The final resulting metric is therefore
\begin{equation}\label{NEM2Metric3}
ds^2 = \frac{1}{q^2} \left( - f(q) dt^2 + d\vec{x}^2_2 + \frac{dq^2}{f(q)} \right),
\end{equation}
which is a special case of the black brane metric \eqref{BBAdSMetric} in $d=3$ dimensions with the horizon at $q_0$. The Hawking temperature of the near-extremal M2 background is therefore given by the black brane expression \eqref{T}. This can verified from the non-extremal temperature where we find that, indeed,    
\begin{equation}\label{M2T}
T = \frac{3}{2\pi r_0} H(r_0)^{- 1/2 }  \to \frac{3 r_0^2}{2\pi},
\end{equation}
in the near-extremal limit. Using the results from section \ref{Sec:AnomDimAdST}, we find the Wilsonian bound
\begin{equation}\label{KRhoxM2}
\frac{\omega^2}{1- \left(2 r_0^2\right)^3 \rhox^3} - \vec{k}^2 \leq \frac{\nu^2 }{ \rho_\times^2} + 18 r_0^6 \rhox,
\end{equation}
where we have used, as before, the variable $\rho$ to indicate the position of the brane along the radial coordinate $q$ in \eqref{NEM2Metric3}. Using $\Lambda = \nu / \rho$ and temperature \eqref{M2T}, we can rewrite \eqref{KRhoxM2} to give the final Wilsonian bound in terms of the undeformed scale and the thermal defomation, $\Lambda^2_{M2} = \Lambda^2 + \Lambda_T^2$:
\begin{equation}\label{KRhoxM22}
\frac{\omega^2}{1- \zeta_1 \left( \frac{T}{\Lambda} \right)^3 } - \vec{k}^2 \leq \Lambda^2 + \zeta_2  \left( \frac{T}{\Lambda} \right)^3 \Lambda^2, 
\end{equation} 
with numerical factors $\zeta_1 = \left( \frac{4 \pi \nu }{3} \right)^3$ and $\zeta_2 =  \frac{16 \pi^3 \nu}{3} $.

\subsubsection{M5 brane}\label{Sec:M5sec}

We can now repeat exactly the same story as in section \ref{Sec:M2sec}, for the M5 brane background. The non-extremal M5 brane metric, without the spherical $d\Omega_4^2$ part is
\begin{equation}\label{M5Metric}
ds^2 = H(r)^{-1/3} \left( -f(r) dt^2 + d\vec{x}^2_5  \right) + H(r)^{2/3} \frac{dr^2}{f(r)},
\end{equation}
where $H(r) = 1 + R^3 / r^3$ and $f(r) = 1 - r_0^3 / r^3$. We use the near-extremal limit $r \ll R =1$, with a different set of coordinates $u^2 = r_0 / r$ in $d=6$ dual boundary field theory dimensions. Hence,
\begin{equation}\label{NEM5Metric}
ds^2 = \frac{r_0}{u^2} \left( - f(u) dt^2 + d\vec{x}^2_5 \right) + \frac{4 du^2}{f(u) u^2},
\end{equation}
with $ f(u) = 1 - u^6 $. We further use $q = \frac{2 u}{\sqrt{r_0}}$ and reabsorb the factor of $4$ into the AdS radius to, again, find a special case of the black brane scenario
\begin{equation}\label{NEM5Metric2}
ds^2 = \frac{1}{q^2} \left( - f(q) dt^2 + d\vec{x}^2_5  + \frac{dq^2}{f(q)} \right),
\end{equation}
where now $f(q) = 1 - \frac{q^6}{q_0^6}$ and $q_0 = \frac{2}{\sqrt{r_0}}$. The Hawking temperature in the near-extremal limit is
\begin{equation}\label{M5T}
T = \frac{3}{4 \pi r_0} H(r_0)^{- 1/2 }  \to \frac{3 r_0^{1/2}}{4\pi}.
\end{equation}
The Wilsonian cut-off in the M5 background is
\begin{equation}\label{KRhoxM5}
\frac{\omega^2}{1- \left( \frac{\sqrt{r_0}}{2} \right)^6 \rhox^6} - \vec{k}^2 \leq \frac{\nu^2 }{ \rho_\times^2} + \frac{9}{64} r_0^3 \rhox^4,
\end{equation}
which we rewrite in terms of the undeformed $\Lambda$ and temperature \eqref{M5T}. Finally, we find
\begin{equation}\label{KRhoxM52}
\frac{\omega^2}{1- \zeta_3 \left( \frac{T}{\Lambda} \right)^6 } - \vec{k}^2 \leq \Lambda^2 + \zeta_4  \left( \frac{T}{\Lambda} \right)^6 \Lambda^2,
\end{equation}    
with $\zeta_3 = \left( \frac{2\pi\nu}{3} \right)^6 $ and $\zeta_4 = \frac{\pi^2}{4} \left( \frac{4\pi\nu}{3} \right)^4  $. All temperature dependence appears, as before, to the power of the boundary field theory dimension $T^d$. Since the M5 and the M2 backgrounds are special cases of the black brane in section \ref{Sec:AnomDimAdST}, all qualitative features of the RG scaling are the same.

\subsection{The general case with mass and thermal deformations}\label{Sec:AnomDimGen}

We now generalise our discussion to a hybrid metric describing the $\CN=4$ theory, or some CFT, with two deformations. The first is a Lorentz symmetry preserving deformation, which can either be a relevant, marginal, or irrelevant term in the action. An example of this was the mass deformation in the GPPZ flow. We denote it by $h(r)$ and refer to it as the mass deformation inducing a mass scale, indicating that it has some general coefficient with a mass dimension. The second, $f(r)$, is a Lorentz symmetry breaking term with a horizon that can describe finite temperature, or finite density effects on the RG flow. We refer to it as the thermal deformation. This means studying a metric $G_{MN}$ of type
\begin{equation}\label{GenMetric}
ds^2 = \frac{h(r)}{r^2} \left( -f(r) dt^2 + d\vec{x}^2_{d-1} \right) + \frac{dr^2}{r^2 f(r)},
\end{equation}
in $d+1$ bulk dimensions, with $h$ and $f$ only depending on the radial coordinate. The next step is to consider equation \eqref{DTRGEq}, with $\lambda_2 = \sqrt{f} (\Delta_- + \gamma(r))$, as dictated by the form of $G^{rr}$. In accordance with previous analyses, we set $\gamma_\times = \nu$ and $\partial_\rho \gamma_\times = 0$ to obtain a Wilsonian bound,
\begin{equation}\label{GenBound}
\frac{\omega^2}{f} - \vec{k}^2 \leq \frac{h}{\rho^2} \left[ \nu^2 - \frac{d^2}{4} \left( 1-f \right) \right]  -  \frac{d}{2 \rho} \left( h \frac{\partial f}{\partial \rho} + \frac{d}{2} f \frac{\partial h}{\partial \rho} \right).
\end{equation}
It is easy to verify that the pure AdS, the GPPZ flow, the massive uncharged black brane, the charged black brane, as well as the M2 and M5 branes are special cases of this expression. To further analyse the Wilsonian bound in \eqref{GenBound}, we assume the form of $f$ with a horizon at $r_0$ to be $f(r) = 1 - \left(r/r_0\right)^p$, and define $h(r) \equiv 1 - \mu^2 (r)$. The Hawking temperature is then
\begin{equation}
T = \frac{ p \sqrt{1-\mu^2(r_0)} }{4 \pi r_0},
\end{equation} 
where $\mu(\rho=r_0)$ only depends on $r_0$ and numerical constants. We can then rewrite \eqref{GenBound} as
\begin{equation}\label{GenBound2}
\frac{\omega^2}{1 - \zeta \left( \frac{T}{\Lambda} \right)^p}  -  \vec{k}^2 \leq \Lambda^2  + \Lambda^2_\CM + \Lambda^2_T - \Lambda^2_{\text{mix}},
\end{equation}
with a constant factor $\zeta =  \left( \frac{4\pi\nu}{pr_0\sqrt{ 1-\mu^2(r_0) } }  \right)^p$, and four different energy scales:
\begin{align}
\Lambda^2	&=	\frac{\nu^2}{\rho^2},\label{GenScale}   \\
\Lambda^2_\CM 	&=	 \left( \frac{d^2}{2\nu^2} \frac{\partial \ln \mu}{\partial \ln \rho} - 1 \right) \Lambda^2 \mu(\Lambda)^2 ,\label{GenScaleM} \\
\Lambda^2_T	&= \left( \frac{dp}{2\nu^2} - \frac{d^2}{4\nu^2} \right)   \zeta   \Lambda^2   \left( \frac{T}{\Lambda} \right)^p,  \label{GenScaleT} 	\\
\Lambda^2_{\text{mix}} &=  \Lambda_\CM^2 \zeta \left( \frac{T}{\Lambda} \right)^p + \mu(\Lambda)^2   \Lambda^2_T+  \Lambda^2 \mu(\Lambda)^2 \zeta \left( \frac{T}{\Lambda} \right)^p . \label{GenScaleMix}
\end{align}
Writing out \eqref{GenScaleMix} explicitly,
\begin{equation}
\Lambda^2_{\text{mix}} =	\left(   \frac{dp}{2\nu^2} - \frac{d^2}{4\nu^2} + \frac{d^2}{2\nu^2}  \frac{\partial \ln \mu}{\partial \ln \rho} \right) \zeta  \Lambda^2  \mu(\Lambda)^2 \left( \frac{T}{\Lambda} \right)^p  .
\end{equation}
In the case of $p=d$, equations \eqref{GenScaleT} and \eqref{GenScaleMix} simplify to give
\begin{align}
\Lambda^2_T	=  \frac{d^2}{4\nu^2}    \zeta   \Lambda^2   \left( \frac{T}{\Lambda} \right)^p  & &\text{and} & &\Lambda^2_{\text{mix}} =  \Lambda_\CM^2 \zeta \left( \frac{T}{\Lambda} \right)^d + \mu(\Lambda)^2   \Lambda^2_T+  \Lambda^2 \mu(\Lambda)^2 \zeta \left( \frac{T}{\Lambda} \right)^d .
\end{align}
It is clear that we can interpret $\Lambda^2$ in \eqref{GenScale} as the undeformed Wilsonian cut-off scale, as in previous sections. $\Lambda^2_\CM$ and $\Lambda^2_T$ (\eqref{GenScaleM}, \eqref{GenScaleT}) are the mass and thermal deformation scales. Depending on the power of $\Lambda$ inside the function $\mu(\Lambda)$, the mass deformation will either modify the UV, or the IR scaling. Lastly, $\Lambda^2_{\text{mix}}$ is a new scale which arises from the mixing of both $\Lambda_T$ and $\Lambda_\CM$. It only exists when both scales are present and is subtracted from the sum of the squares of other three scales, giving us the total Wilsonian cut-off of the deformed theory. 

Using the results in \eqref{GenScale}-\eqref{GenScaleMix}, we can generalise our discussion to include an arbitrary number of mass and thermal deformations. We simply write $h$ and $f$ as a product,
\begin{align}
h(r) = \prod_{i=1}^n h_i(r)  & &\text{and}& &  f(r) = \prod_{j=1}^m f_j(r),
\end{align}
giving
\begin{align}\label{GenBound3}
\frac{\omega^2}{\prod_j f_j} - \vec{k}^2 \leq \frac{\prod_i h_i}{\rho^2} \left[ \nu^2 - \frac{d^2}{4} \left( 1-\prod_j f_j \right) \right]  -  \frac{d }{2 \rho} \left( \prod_{i,j} h_i f_j \right) \left( \sum_j \frac{\partial \ln f_j}{\partial \rho} + \frac{d}{2} \sum_i \frac{\partial \ln h_i}{\partial \rho} \right).
\end{align}
Imagine now that we have a sequence of horizons at $r_1 < r_2 < ... < r_m$, so that in flowing from $r=0$ we cannot run past $r_1$. Writing $f_j = 1 - (r/r_j)^{p_j}$, and expanding around $u^2 = \alpha (r - r_1)$ gives the Hawking temperature, 
\begin{align}
T =  \frac{p_1 \sqrt{h(r_1)} \prod_{j=2}^{m}f_j (r_1)}{4\pi r_1},
\end{align}
which can then be used to rewrite \eqref{GenBound3} in terms of temperature $T(r_1)$ and the undeformed scale $\Lambda = \nu / \rho$, which always takes the same form. 

Lastly, it is essential to note that in order to have a well-defined, physical RG flow, the scale $\Lambda = \nu / \rho$ has to be real and positive. This condition is therefore always equivalent to the Breitenlohner-Freedman bound whereby $\nu$ also has to be real and positive. This shows that our construction of the Wilsonian RG is consistent with the AdS/CFT dictionary, as well as that the results are clearly probe-dependent.

\section{Alternative interpretation of the double-trace flow and the c-theorem}\label{Sec:Alter}

In this section, we comment on alternative interpretations of the renormalisation group flow we have been analysing so far. We begin by returning to the structure of the holographic counter-terms in equation \eqref{CTBdyAct}. In the limit of $\rho_0 \to 0$, we could neglect all terms proportional to the conformal anomaly, $\ln (\rho)  \Phi \Box_g \Phi$, and cancel terms proportional to the Ricci curvature of the induced flat boundary metric, $R[g] \Phi^2$ \cite{de Haro:2000xn}. Notice that both terms appear with quadratic powers of $\Phi$. We could therefore reinterpret the running of $(\Delta_- + \gamma) \Phi^2$ as either one by absorbing them into $\gamma$. 

Firstly, imagine that we keep our fixed asymptotically AdS metrics with flat boundary foliations, so that $R[g]=0$. The conformal anomaly only appears in an even number of boundary dimensions $d$. In that case, we reinterpret $\gamma$ as the coefficient in front of the conformal anomaly, which will be proportional to the expectation value of the boundary stress-energy tensor $\langle T^\mu_\mu \rangle$. Hence $\gamma$ can be interpreted as being proportional to the c-function. We saw in all examples that $\gamma$ was a monotonic function, which is therefore consistent with the c-theorem \cite{Zam:1986,Cardy:1988,Freedman:1999gp,Komargodski:2011vj,Myers:2010xs} . 

Imagine instead that we still have an asymptotically AdS bulk, but with the dual boundary theory living on a $d$-dimensional sphere, $S^d$. Its Ricci curvature, $R[g] = d (d-1) / \ell^2$ is inversely proportional to the square of its radius $\ell$. At the AdS infinity, $\ell$ also goes to infinity so $R[g]$ vanishes. When we start flowing into the bulk, the running of $\gamma$ can now be interpreted as the running of the radius $\ell$. And since $\gamma$ monotonically increases, the radius $\ell$ monotonically decreases. It is a convenient property of CFT's on $S^d$ that their central charge can be defined by the integral $c = \left\langle \int_{S^d} d^d x \sqrt{-g} T^\mu_\mu \right\rangle$ \cite{Gubser:2002vv,Hartman:2006dy}. We can match the calculated function $\gamma$ with a monotonically increasing $R[g]$ and then use the Einstein's equation to recover the flow of $T^\mu_\mu$. This implies that the c-function can easily be determined throughout the flow where it remains monotonic. The analysis is therefore again consistent with the c-theorem.

\section{The effective action with multi-trace deformations}\label{Sec:MultiTrace}

We now return to the effective Wilsonian action \eqref{BareActRho} with a full set of multi-trace deformations. We expect the effective action to include all possible terms consistent with the symmetries of the theory. An infinite series of multi-trace deformations should therefore run under the RG flow, unless a reduced form of the effective action closes in on itself under successive integrations. In case of a Gaussian action, for example, only double-trace deformations run. Even though, as argued in section \ref{Sec:RG3}, all triple-and higher-trace deformations are subleading in the large-$N$, they generally turn on and contribute to the flow, in analogy with the Wilsonian $\epsilon$-expansion. We could safely neglect them in extracting the functional dependence of the cut-off on the bulk, but it is important to analyse their behaviour for further establishing the Wilsonian nature of the RG procedure under consideration.  

Given the discussion above, we make the following claim: \textit{the series of running terms in the effective action \eqref{BareActRho} can either be quadratic or infinite}. The only other cases could emerge when there is some special finely-tuned relation between the metric and the scalar potential, coming from some symmetry of the uncompactified, supergravity theory. We will now show that this is precisely what happens in our holographic construction of the Wilsonian RG, as governed by equation \eqref{HFlowEq2}, and hence \eqref{PiRGeq}, \eqref{Coeff2RGeq}, \eqref{CoeffnRGeq} and \eqref{AlphaRGeq}. Equation \eqref{PiRGeq} was found to describe the Callan-Symanzik equation on the QFT side. Equation \eqref{Coeff2RGeq} describes the flow of the anomalous dimension and, equivalently, the double-trace beta function. Furthermore, \eqref{CoeffnRGeq} describes different multi-trace beta functions. Flows of different multi-trace couplings depend on each other, which is reflected in the system of coupled \eqref{Coeff2RGeq} and \eqref{CoeffnRGeq}.

To prove this statement, let us assume that the series of terms $\lambda_n \Phi^n$ terminates at some $N$ in the effective action, such that $\lambda_{m} = 0$, for all $m \geq N+1$. We are left with a finite number of coupled differential equations \eqref{PiRGeq}, \eqref{Coeff2RGeq} and \eqref{CoeffnRGeq}, which we refer to by the $n^{th}$-order coefficient they describe. The flow of $\Pi$ is described by the $1^{st}$-order \eqref{PiRGeq}. We will not use the $0^{th}$-order equation \eqref{AlphaRGeq}, which gives the cosmological constant. The left-hand side of the Hamiltonian RG flow equation \eqref{HFlowEq2} includes terms up to $\partial_\rho \left(\sqrt{-g} \lambda_N \Phi^N\right)$. The right-hand side, however, produces terms up to the order of $\Phi^{2N - 2}$, resulting from $\left( \delta S_B / \delta \Phi \right)^2$. Matching terms in \eqref{HFlowEq2}, order-by-order in $\Phi$, thus gives $2N-2$ equations, disregarding the $0^{th}$-order flow of $\alpha$. The first $N$ are differential equations for $\lambda_n$, i.e. \eqref{CoeffnRGeq}, whereas the remaining $N-2$ are algebraic and relate the non-zero $\lambda_{n\leq N}$ among themselves. They are
\begin{equation}\label{AlgEqMT}
0 = - \frac{n}{2} \sum_{m=2}^{n} \lambda_m \lambda_{n+2-m}  + b_n,~\text{for}~N+1\leq n \leq 2N-2.
\end{equation}  
We can immediately see that all couplings in the scalar potential, with $n > 2N - 2$, must be equal to zero, $b_{n > 2N-2} = 0$. To study the constraints that equations \eqref{AlgEqMT} impose on $\lambda_n$, we can recursively solve the system of algebraic equations by starting with $n=2N-2$. Since all $\lambda_{m \geq N+1} = 0$, terms $\lambda_2 \lambda_{2N-2} + \lambda_3 \lambda_{2N-3} + ... +\lambda_{N-1}\lambda_{N+1} $ vanish, leaving only $\lambda^2_N$ in the sum. Therefore,
\begin{equation}\label{LNasB}
\lambda_N^2 = \frac{b_{2N-2}}{N-1},
\end{equation}
which completely fixes the value of $\lambda_N$ by the last coupling in the potential, $b_{2N-2}$. At the order of $n=2N-3$, equation \eqref{AlgEqMT} has a series $\lambda_2 \lambda_{2N-3} + ... + \lambda_{N-2} \lambda_{N+1}$ of vanishing terms and a non-zero $\lambda_{N-1} \lambda_N$. We find
$\lambda_{N-1} \lambda_N = \frac{b_{2N-3}}{2N-3}$, which implies 
\begin{equation}
\lambda_{N-1} = \frac{b_{2N-3}}{2N-3} \sqrt{  \frac{N-1}{b_{2N-2}}}.
\end{equation}
The coefficient $\lambda_{N-1}$ is completely fixed by the two couplings $b_{2N-3}$ and $b_{2N-2}$. It is easy to see that this behaviour recursively continues as we solve for other $\lambda_n$. At $n=2N-4$, we have non-zero $2 \lambda_{N-2} \lambda_N +  \lambda_{N-1}^2$. Similarly, at each step in the recursive process, the undetermined, lowest order $\lambda_{N-l}$ couples only to $\lambda_N$. All other $\{\lambda_{N-l+1}, ... , \lambda_N\}$ are at that step already fixed by $\{b_{2N-2-l+1}, b_{2N-2-l+2}, ... ,b_{2N-2}\}$. We can, therefore, solve for $\lambda_{N-l}$ in terms of $b_{n\geq 2N-2-l+1}$ and $b_{2N-2-l}$, appearing in \eqref{AlgEqMT} at the $(2N-2-l)^{th}$ order. Hence, all $\lambda_{N-l}$ are completely fixed by the set of couplings, $\{b_{2N-2-l}, b_{2N-2-l+1},...,b_{2N-2}\}$. Now, since we have $N-2$ equations \eqref{AlgEqMT}, this process ends at $l=N-3$ with fixing $\lambda_3$ in terms of $\{b_{N+1}, b_{N+2},...,b_{2N-2} \}$. At the lowest order of \eqref{AlgEqMT}, at $n=N+1$, there is the term $\lambda_2 \lambda_{N+1} = 0$, leaving $\lambda_2$ undetermined. The anomalous operator dimension $\gamma$ and, equivalently, the double-trace coupling are therefore not constrained by the algebraic equations \eqref{AlgEqMT}.

Coupling constants $b_n$ do not depend on $\rho$ as they are the coefficients of the classical scalar potential. If we worked with the full quantum field theory in the bulk, it is possible that they would receive renormalisation group contributions which could depend on $\rho$. We, however, keep the bulk theory classical in correspondence with the large-$N$ boundary theory. Similarly, $b_n$ cannot depend on momentum $k$. We have so far shown that the assumption of a finite series $\lambda_n \Phi^n$ in the effective action implies that all $\lambda_{3\leq n\leq N}$ are uniquely determined by $b_{ N+1 \leq n \leq 2N-2}$. They are therefore constant for all $3\leq n\leq N$, $\partial_\rho \lambda_{n} = 0$, and equal the initial values, $\lambda_{n} = a_{n}$, which are set by the holographic counter-terms. Equation \eqref{CoeffnRGeq} now simplifies to give
\begin{equation}\label{DiffEqMT}
\frac{\partial_\rho \sqrt{-g(\rho)} }{\sqrt{-G(\rho)}} \lambda_n = - n \lambda_2 (\rho)\lambda_n  - \frac{n}{2} \sum_{m=3}^{n-1} \lambda_m \lambda_{n+2-m} + n \lambda_{n+1} \frac{\Pi(\rho)}{\sqrt{-g(\rho)}} + b_n, ~\text{for}~  3 \leq n \leq N.
\end{equation} 
When $n=N$, there is no $\lambda_{N+1} \Pi$ term, as $\lambda_{N+1} = 0$. Now, since $b_n$ are $k$-independent, so are $\lambda_n$, $\partial_k b_n = \partial_k \lambda_n = 0$, for $n\geq 3$. We should note that this depends on the fact that our probe scalar action included no higher derivatives than second in the kinetic term. If it had, the multi-trace RG equations would have to include various momentum terms in momentum space. The bulk metric $G_{MN}$ is also independent of the scalar momentum $k$, hence $\partial_k G = \partial_k g = 0$. The only momentum dependence in \eqref{DiffEqMT} therefore comes from $\lambda_2$ and $\Pi$, which both have to depend on $k$ due to \eqref{PiRGeq} and \eqref{Coeff2RGeq}. Differentiating \eqref{DiffEqMT} with respect to $k$ gives
\begin{equation}\label{DiffEqMT2}
\lambda_n \frac{\partial \lambda_2}{\partial k} = \lambda_{n+1} \frac{1}{\sqrt{-g}} \frac{\partial \Pi}{\partial k}, ~\text{for} ~3\leq n \leq N.
\end{equation}
For $n=N$ with $\lambda_{N+1} = 0$, \eqref{DiffEqMT2} immediately gives $\lambda_N = 0$. Using \eqref{LNasB} then implies that $b_{2N-2} = 0$. But now since $\lambda_N = 0$, equation \eqref{DiffEqMT2} implies that $\lambda_{N-1} = 0$. We can therefore recursively see from \eqref{DiffEqMT2} that $\lambda_{N} = 0 \Rightarrow \lambda_{N-1} = 0 \Rightarrow \lambda_{N-2}  = 0 \Rightarrow ... \Rightarrow \lambda_4 = 0 \Rightarrow \lambda_3 = 0$. As a result of that, all $b_n = 0$, for $N+1 \leq n \leq 2N-2$. Furthermore, from the full \eqref{DiffEqMT2} we can see that all $b_n = 0$, for $3 \geq n \geq N$, as well. Hence we have shown that, assuming $S_\text{B}$ had a finite number of terms $\lambda_n \Phi^n$, the RG flow equations enforce that \textit{all} $\lambda_n$ and $b_n$, for $n \geq 3$, \textit{be zero throughout the flow}. The only non-zero terms left are those that we studied in the context of the leading order in the large-$N$, i.e. the double-trace sector: $\Pi$, $\lambda_2$ and $m$. We have, therefore, shown that indeed, as claimed, the Wilsonian effective bare boundary action can either be quadratic or infinite. Only the double-trace sector can close in on itself under the RG procedure, as is expected in the Wilsonian RG. This fact is also in agreement with the semiclassical path integral derivation of the Wilsonian renormalisation from holography in the work of \cite{Heemskerk:2010hk}. Lastly, the double-trace sector, with $N=2$, is also the only case where the number of equations describing the RG flow matches the number of running terms, i.e. $2N-2=N=2$.

From our discussion, we see that only in the case when we are working with the double-trace sector is it allowed to terminate the series of terms in the effective action. Otherwise we must include all possible multi-trace terms, even if the holographic counter-terms vanish for certain powers of $\Phi$. We should still allow all terms to run and use the vanishing initial conditions for the flow of the zero holographic counter-terms. It is still possible that then some of them will remain zero throughout the flow. As argued before, additional symmetries could restrict the form of multi-trace couplings and make some of them vanish. Our analysis therefore confirms the Wilsonian nature of the renormalisation procedure we set up from holography. The full, infinite set of coupled differential equations which describe the Wilsonian RG flow is, however, in general extremely hard to solve in the absence of special symmetries in the problem. We could, in principle, recursively look for a solution in the following way, through purely algebraic manipulations. First express $\lambda_2$ as $\Pi$ by using the $1^{st}$-order equation \eqref{PiRGeq}, giving us $\lambda_2 = - \frac{\sqrt{G^{rr}}}{2} \frac{\partial \ln \Pi}{\partial \rho}$. Then insert $\lambda_2 (\Pi)$ into the $2^{nd}$-order equation \eqref{Coeff2RGeq}, to obtain $\lambda_3 (\Pi)$. At the $3^{rd}$-order in \eqref{CoeffnRGeq}, we use both $\lambda_2 (\Pi)$ and $\lambda_3 (\Pi)$ to recover $\lambda_4(\Pi)$, and then continue with the same procedure. At each recursive step we can express the next $\lambda_n$ in terms of $\Pi$, $G_{MN}$ and their derivatives. This process then continues infinitely many times and we can hope to uncover some symmetry or a systematic progression between these extremely complicated expressions, with growing complexity at each higher-order of $\lambda_n$. 

There is an interesting and powerful simplification of the renormalisation group equations that we now consider as the final part of the multi-trace analysis. We begin with the argument from section \ref{Sec:RG3}, that all multi-trace interactions higher than double-trace, are subleading in the large-$N$ limit. As a result, we could neglect the $\lambda_3 \Pi$ term in the double-trace equation \eqref{Coeff2RGeq} at the leading order in $N$. Note, of course, that $N$ in this final discussion means the number of the gauge group colours, or matrix elements, and not the $N$ that we have so far been using to indicate the $N^{th}$-order multi-trace coupling where we terminated the effective action series. Now, we can similarly imagine a generalisation of this argument to a scenario where the $(n+1)$-point vertices in Feynman diagrams, which are proportional to $\lambda_{n+1}$, contribute only at a subleading order to the computation of the $\lambda_n$ beta function. We can therefore recover a partial re-summation of the diagrams that contribute to the beta function of $\lambda_n$ by neglecting $\lambda_{n+1} \Pi$ in \eqref{CoeffnRGeq}. We get
\begin{equation}\label{MTlargeN}
\frac{1}{\sqrt{-G}} \partial_\rho \left( \sqrt{-g} \lambda_n \right) = - n \lambda_2 \lambda_n - F_n \left[\lambda_3,...,\lambda_{n-1}\right]  + b_n,
\end{equation}
where $F_n \left[\lambda_3,...,\lambda_{n-1}\right]  = \frac{n}{2} \sum_{m=3}^{n-1} \lambda_m \lambda_{n+2-m}$, and is independent of $\lambda_n$. These first-order differential equation can be analytically solved through recursive integration, at least formally, for all multi-trace couplings $\lambda_n$. The solutions of \eqref{MTlargeN} are
\begin{align}\label{MTnSol}
\lambda_n = a_n e^{- \!\! \int\limits_{r=\rho_0}^{r=\rho} \!\!\! L_n(r) d\ln r} \!\! +  e^{- \!\! \int\limits_{r=1}^{r=\rho} \!\! L_n(r) d\ln r} \!\! \int\limits_{r=\rho_0}^{r=\rho} \!\!  \left( b_n - F_n(r)\right) \exp\left\{ \;\; \int\limits_{r'=1}^{r'=r} \!\!\! L_n(r') d\ln r'\right\}  d\ln r ,
\end{align}
where we have defined
\begin{equation}
L_n [\lambda_2, g] (\rho)  \equiv n \lambda_2 (\rho) + \frac{1}{2}\frac{\partial \ln g(\rho)}{\partial \ln \rho},
\end{equation}
and used the Anti-de Sitter-like $G_{rr} = 1 / r^2$ to simplify the form of the solution. We could have also solved \eqref{MTlargeN} for a general function $G_{rr}$, without any additional complications. At the $n^{th}$-order, all $\{\lambda_2,...,\lambda_{n-1}\}$ are already known, so we can insert these functions in solution \eqref{MTnSol}, and integrate to obtain $\lambda_n$. We used $\rho_0$ to indicate the brane position where we specify the boundary conditions, $\lambda_n = a_n$. Due to our construction, we can then take $\rho_0 \to 0$ and obtain finite, well-defined results for the recursive solution of the full set of running multi-trace couplings.

\section{Summary and outlook}\label{Sec:Sum}

In this paper, we proposed a systematic procedure for finding a precise functional dependence of the Wilsonian cut-off scale, in theories with scalar operators, on the bulk quantities. This procedure can be applied to a wide range of theories. To establish the relation between the bulk and the field theory cut-off scale, we constructed the effective running bare boundary action from the structure of the holographic counter-terms. Specifically, using a combination of the counter-terms and the subtracted action, from which the renormalised correlators are calculated, we rewrote the bare action and allowed for various terms in it to run. Since the counter-terms must always transform under the bulk isometries, the structure is consistent with the expectation that all additional terms in the Wilsonian effective action must be invariant under the symmetries of the original theory. The isometries of the bulk are in fact dual to the symmetries of the boundary QFT. The construction depended crucially on working with asymptotically AdS spacetimes, as the counter-term structure simplifies significantly in the AdS limit. We only focused on scalar bulk actions with dual scalar operators. However, this construction can be applied to any other theory once all the relevant holographic counter-terms are known. Of particular interest are the bulk fields with higher spins, which are dual to anomalous globally conserved boundary operators. Operators without anomalous symmetries are not expected to run under RG transformations. Interesting examples of such anomalous operators are the stress-energy tensor dual to the bulk graviton, which runs as a result of the conformal anomaly, and the R-symmetry current dual to the bulk vector field, which runs due to the triangle anomaly. A detailed analysis of the Wilsonian cut-off scale dependence on bulks with various types of bulk fields was not the main purpose of this paper. We leave that work, in particular the case of dynamical gravity, for future investigation. 

The renormalisation group equations were derived by integrating out thin slabs of geometry, as in \cite{Faulkner:2010jy}, starting from the AdS infinity. The Hamilton-Jacobi equation in the bulk, with the radial direction replacing time, then evolved the effective action. Using the standard (Dirichlet) quantisation, where the conjugate canonical momentum is proportional to the expectation value of the boundary operator $\langle \CO \rangle$, we made connection with the Callan-Symanzik equation of the two-point function. We introduced the wavefunction renormalisation $Z$, which transformed under the sliding of the brane, $\rho \to \rho + \delta \rho$, and induced a rescaling transformation on the QFT side of the duality. By noting that fully renormalised two-point functions must be invariant under the RG transformations, we could connect $Z$ with the anomalous operator dimension $\gamma$, coming from the running of the counter-term $\Delta_- \Phi^2$. The RG flow of the anomalous dimension was shown to coincide with the flow of the double-trace deformation, as expected in large-$N$ theories.  

We imposed renormalisation conditions at the operator scale $\mu = \sqrt{-k^2}$, as well as physical conditions on the behaviour of anomalous dimensions, such as reality, non-singularity and their having to obey the unitarity bound. The first renormalisation condition was the cut-off independence of the anomalous dimension, i.e. $\partial_\rho \gamma = 0$ at the RG scale $\mu = \sqrt{-k^2}$. The second condition set the value of the anomalous dimension to $\gamma = \nu$ when the running cut-off $\Lambda$ reached the RG scale. It followed directly from our imposing the following condition on the value of a bare two-point function at the observational RG scale. Namely, since no IR divergences were present in our theories, we demanded that the bare correlation function should equal its initial value with the cut-off $\Lambda_0$ in the extreme UV, where holographic renormalisation is performed. Equivalently, the overall rescaling wavefunction renormalisation of $\CO$, where the flow terminates, had to be equal to one. We then considered several examples, and a precise correspondence between the cut-off and the bulk was established. It is interesting to note that the hard Wilsonian cut-off was Lorentzian, contrary to the usual field theory examples. We were therefore effectively integrating out momenta below a running hyperbola in a light-cone diagram. And although such a cut-off may not be sufficient for removing infinities from loop integrals, it makes perfect physical sense. It is completely consistent with the expectation that there should exist a relativistic ordering of physical phenomena according to their invariant length scales. The Euclidean IR/UV hierarchy with a separation of energy and momentum then mixes and a new relativistic hierarchy may emerge.    

In the simplest scenario of the $\CN = 4$ theory with a pure AdS bulk dual, we confirmed that $\Lambda \propto 1 / r$. Our construction also allowed us to find the precise probe-dependent proportionality constant, $\nu$, which could be changed by coordinate transformations in the bulk. This nicely shows how details of the boundary theory depend on the bulk coordinates. It also speaks to the fact that our results are probe-dependent and future work will be necessary to attempt to find a complete probe-independent Wilsonian description of the dual QFT. Next, we added an IR mass deformation that broke the theory down to $\CN=1$, i.e. the GPPZ flow. We saw that the square of the overall Wilsonian cut-off equaled $\Lambda^2 + \Lambda_\CM^2$, where $\Lambda$ was the undeformed Wilsonian scale and $\Lambda_\CM$ the mass deformation scale. We could also see the existence of a mass gap at the end of the flow. In the case of $\CN=4$ at finite temperature, similarly, the Wilsonian cut-off equaled $\Lambda^2 + \Lambda_T^2$, where the thermal scale $\Lambda^2_T$ depended on $(T/\Lambda)^d$ and $\Lambda^2$. In this theory, a mass gap emerged at the horizon. An interesting feature we found was that the thermal scale in $d=2$ boundary dimensions, where it was independent of $\Lambda$ (and therefore $\rho$), equaled $\Lambda^2_T$ in any number of dimensions when the boundary was pushed to the horizon. This suggests a dimensional reduction of the effective IR CFT scaling reminiscent of the studies into the emergent two-dimensional Virasoro algebras of asymptotic symmetries describing black hole horizons, \cite{hep-th/9812013,hep-th/9812056}. We compared the thermal case with a charged massive black brane, which gives rise to finite temperature and density of the boundary QFT. We found that at the horizon, the Wilsonian scale involved a mixing of two temperatures - one at zero and another at non-zero density of the system. In two dimensions, the deformation scale equaled the purely thermal scale from the example at zero density. It, however, did not exactly match the deformation scale at the horizon in arbitrary $d$. Further study will be required to understand the physics of these IR CFTs. We also analysed the near-extremal M2 and M5 brane backgrounds, which are special cases of the massive $d$-dimensional black brane. All results in the two M-theory backgrounds could therefore be directly obtained from the black brane analysis.  

Finally, we generalised our discussion to a metric with both a horizon, inducing Lorentz violating temperature or density, and a Lorentz invariant deformation. The latter could be a relevant, marginal or irrelevant term with some mass dimension of the coupling. We saw that the overall Wilsonian cut-off equaled to the sum of the squares of the three scales, minus a mixing term which only existed when there were two or more deformations in the theory. We also showed how our generalised discussion could be extended to include an arbitrary number of mass deformations and horizons. It would be particularly interesting to further look at cases where horizons are represented by smooth functions and hence dual to smooth scales. The flow could then pass through them. We leave this problem for future investigation. At the end of the section we commented on the fact that for our construction to be well-defined and physical the running undeformed scale had to be real and positive. This consistency condition turned out to be precisely the Breitenlohner-Freedman bound on the scalar probe masses.

We then went on to briefly comment on how the flow of the anomalous dimension (or the double-trace deformation) could be reinterpreted as the flow of the conformal anomaly, or alternatively the Ricci curvature of the boundary manifold. In our example, we used the $S^d$ manifold. Both cases were consistent with the c-theorem, which was ensured by the monotonicity of $\gamma$ throughout the flow. Further study, in particular of various asymptotically AdS manifolds with curved boundary foliations is kept for future work.
  
In the last section, we studied the constructed Wilsonian effective bare boundary action with a full, infinite set of multi-trace terms. In order to provide further evidence beyond the existence of a momentum cut-off that the renormalisation group procedure under consideration is truly Wilsonian, we showed that the effective action must either be quadratic (double-trace) or have an infinite series of multi-trace terms. This was shown by analysing the full set of the renormalisation group equations, which we had derived from holography: the Callan-Symanzik equation, the flow of the anomalous dimension and the double-trace beta function, and the infinite set of multi-trace beta functions. The assumption that the bare boundary action $S_\text{B}$ only included a finite number of terms automatically lead us to the conclusion that all multi-trace couplings $\lambda_{n}$, for $n\geq 3$, had to be equal to zero throughout the entire RG flow. As a result, all coupling constants in the scalar potential also had to be zero, with the exception of mass. We therefore showed that only the double-trace sector could close in on itself under successive integrations, as in a Gaussian field theory. Otherwise, in the absence of special finely-tuned symmetries, an infinite series of all possible multi-trace terms permitted by symmetries had to be present in the effective action. These results are therefore completely consistent with the Wilsonian renormalisation group in quantum field theory. Lastly, we commented on possible recursive procedures for obtaining analytic solutions to the full set of the multi-trace flows.


\acknowledgments{
The author would like to thank Andrei Starinets for numerous insightful discussions, encouragement, as well as for comments on a draft of this manuscript. The author is thankful to Janos Polonyi and Mukund Rangamani for extremely useful discussions and comments on the manuscript. The author is also grateful to Edward Hardy, Nikolaos Kaplis, Joan Simon and Eirik Svarnes for helpful discussions. S.~G. is supported by the Graduate Scholarship of St. John's College, Oxford.}


\end{document}